\documentclass[psfig,useAMS,usenatbib]{mn2e}
\usepackage{times}
\usepackage{epsfig}
\usepackage{amsmath}
\usepackage{natbib}
\usepackage{longtable}
\def\mpc{h^{-1}{\rm{Mpc}}}
\def\kms {\rm{km~s^{-1}}}
\def\apj {ApJ}

\def\apjs {ApJS}
\def\aj {AJ}
\def\azh {Astronomicheskii Zhurna}

\def\mnras {MNRAS}

\def\sovast {Sov. Astron.}

\def\zphot {z_{\rm phot}}
\def\zspec {z_{\rm spec}}

\def\mpc{{\  h^{-1} \rm Mpc}}
\def\kpc{{\ h^{-1} \ \rm kpc}}
\def\kms{{\rm km \ s^{-1}}}
\title[Galaxy triplets in SDSS-DR7.]
      {Galaxy triplets in SDSS-DR7: I. Catalogue}
\author[O'Mill et al.]
  {Ana Laura O'Mill$^{3}$\thanks{E-mail: aomill@oac.uncor.edu}, 
Fernanda Duplancic$^{4,2}$, Diego Garc\'\i a Lambas$^{1,2}$, 
Carlos Valotto$^{1,2}$ \& \newauthor Laerte Sodr\'e Jr$^{3}$ \\ 
  $^1$ Instituto de  Astronom\'\i a Te\'orica y Experimental, IATE, Observatorio Astron\'omico, Universidad Nacional de C\'ordoba,\\ 
Laprida 854, X5000BGR, C\'ordoba Argentina\\
  $^2$Consejo de Investigaciones Cient\'\i ficas y T\'ecnicas (CONICET),\\Avenida Rivadavia 
1917, C1033AAJ, Buenos Aires, Argentina\\
  $^3$ Departamento de Astronomia, Instituto de Astronomia, Geof\'\i sica
e Ci\^encias Atmosf\'ericas da USP,\\ Rua do Mat\~ao 1226, Cidade
Universit\'aria, 05508-090, S\~ao Paulo, Brazil.\\
  $^4$ Instituto de Ciencias Astron\'omicas, de la Tierra y del Espacio, ICATE, Casilla de Correo 49, CP 5400, San Juan, Argentina.}
\pagerange{\pageref{firstpage}--\pageref{lastpage}} \pubyear{2002}
\def\LaTeX{L\kern-.36em\raise.3ex\hbox{a}\kern-.15em
    T\kern-.1667em\lower.7ex\hbox{E}\kern-.125emX}

\begin{document}
\label{firstpage}
\maketitle

\begin{abstract}

We present a new catalogue of galaxy triplets derived from the Sloan Digital
Sky Survey Data Release 7. The identification of systems was performed 
considering galaxies brighter than $M_r=-20.5$ and imposing constraints over 
the projected distances, radial velocity differences of neighbouring galaxies
and isolation. To improve the identification of triplets we employed a data 
pixelization scheme, which allows to handle large amounts of data as in the 
SDSS photometric survey.
Using spectroscopic and photometric data in the redshift range $0.01\le z\le 0.40$ 
we obtain $5901$ triplet candidates. We have used  a mock catalogue to analyse the completeness and 
contamination of our methods. The results show a high level of completeness ($\sim 80 \%$) 
and low contamination ($\sim 5 \%$). By using photometric and spectroscopic data we have also 
addressed the effects of fiber collisions in the spectroscopic sample.
We have defined an isolation criterion considering the distance of the
triplet brightest galaxy to closest neighbour cluster, to describe a global 
environment, as well as the galaxies within a fixed aperture, around the
triplet brightest galaxy, to measure the local environment. The final catalogue comprises $1092$ isolated 
triplets of galaxies in the redshift range $ 0.01\le z\le 0.40$.
Our results show that photometric redshifts provide very useful 
information, allowing to complete the sample of nearby systems whose detection
is affected by fiber collisions,  as well as extending the detection of 
triplets to large distances, where spectroscopic redshifts are not available.

\end{abstract}

\begin{keywords}
Galaxies: systems: general
\end{keywords}


\section{Introduction}

In a hierarchical scenario, the interactions affects general properties of the galaxies such as star 
formation rate, nuclear activity and morphology \citep{alonso04, alonso06}

Although galaxy pairs \citep{Lambas03, alonso04, alonso06} and groups with four or 
more members \citep{manuel} have been the subject of several papers in the literature, 
there are far few works on triplets of galaxies, a link between pairs and 
groups and a potentially interesting laboratory to investigate several process 
associated to galaxy evolution. 

Traditionally, galaxy groups selection have excluded triplets of 
galaxies because these systems were not considered as bonafide physical 
structures, i.e., dynamically isolated systems. 
Nevertheless,  \citet{HT11} have recently analysed a sample of galaxy triplets 
derived from the catalogue of isolated triplets of \citet{kara73} finding
signatures of physical interactions (tidal bridges, excess of barred galaxies),
that implies that most of these objects are indeed real physical structures.

Pioneering works on the identification of triplets of galaxies were 
performed by \citet{karachen} and \citet{karachen88}. 
These authors present a list of 84 Northern isolated galaxy 
triplets with members with component apparent magnitudes brighter than 15.7, 
selected by visual inspection of Palomar Sky Survey prints. 
They found that about 24$\%$ of the members are elliptical and lenticular 
galaxies, while 76$\%$ are spirals and irregulars. Numerous studies have been made in the 
follow up of this catalogue. \citet{karachen81} measured radial velocities for the 
isolated triplets and found that 5$\%$ of the systems are spurious ($\Delta V_{ij}>500 \kms$), 
31$\%$ consist of a pair of galaxies with nearly equal radial velocities ($\Delta V_{ij}<500 \kms$) 
and a projected third component and 64$\%$ are considered physical triplets 
($\Delta V_{ij}<500 \kms$). Velocity dispersion, diameters, integrated luminosity, virial mass, 
and the virial mass to luminosity ratio were estimated for the physical 
triplets \citep{karachen82,karachen83}. The spatial configuration, dynamics and 
dark matter content of these systems was studied by several authors 
\citep{karachen90, chernin91, anosova, zheng93, aceves}. For the Southern sky 
($\delta<-3\textordmasculine$) \citet{karachen2000} selected 76 isolated triple 
systems of galaxies using the ESO/SERC and POSS-I sky surveys, with the same criteria 
defined fort the Northern systems.

\citet{Wide} compiled a list of 108 triplets of galaxies selected from 
two group catalogues, those of  \citet{GH} (Northern sky) and \citet{Maia} 
(Southern sky). 38 of these systems are considered isolated. The authors call these systems 
``wide triplets'' in contrast to the compact triplets (with smaller mean projected harmonic separation) 
studied by \citet{karachen}. The main assumed difference between wide and compact systems, 
regardless their difference in size, is that galaxies in compact triplets are
supposed to have interacted many times and are probably in equilibrium within the system. In contrast, wide 
triplets would be  dynamically young and probably far from virialization.

A high-order 3D Voronoi tessellation method was employed by \citet{elyiv09} 
for identifying single galaxies, pairs and triplets on a volume-limited sample of galaxies from 
the Sloan Digital Sky Survey Data Release 5. Nevertheless, a catalogue of triplet of galaxies 
has not been compiled for more recent SDSS releases. 

In this work we present a catalogue of triplets of bright galaxies ($M_r<-20.5$) 
selected from a volume limited sample extracted from SDSS-DR7 in the redshift range $0.01\le z\le 0.4$. We used 
data with both spectroscopic redshift measurements and photometric 
redshifts from \citet{photo}.

This paper is organized as follows. In Section 2 we describe the galaxy samples used in this work. Section 3 
describes the algorithm developed for the detection of triplet of galaxies candidates. 
The implementation of this algorithm at low redshifts is described in Section 4. 
In this section we also perform a completeness and contamination test to our algorithm and analyse 
the incompleteness effect due to fiber collisions. Section 5 describes the extension of our results to higher redshifts. 
In Section 6 we describe the isolation criteria employed to define the final sample of 
triplet of galaxies. Finally, Section 7 summarized the results obtained in this work.

Throughout this paper we adopt a cosmological model characterised by
the parameters $\Omega_m=0.25$, $\Omega_{\Lambda}=0.75$ and $H_0=70{\rm km~s^{-1}~Mpc^{-1}}$.


\section{Spectroscopic and photometric galaxy samples}

The samples of galaxies used in this work were drawn from the Data Release 7 
of Sloan Digital Sky Survey  \citep[SDSS-DR7,][]{dr7}.
SDSS \citep{sdss} has mapped more than one-quarter of the entire sky, 
performing  photometry and spectroscopy for galaxies, quasars and 
stars. SDSS-DR7 is the seventh major data release, corresponding to the
completion of the survey SDSS-II. It comprises $11.663$ sq. deg.
of imaging data, with an increment of $\sim2000$ sq. deg., over the 
previous data release, mostly in regions of low Galactic latitude.
SDSS-DR7 provides imaging data for 357 million 
distinct objects in five bands, \textit{ugriz}, as well as
spectroscopy  over $\simeq \pi$ steradians in the North Galactic 
cap and $250$ square degrees in the South Galactic cap. 
The average wavelengths corresponding to the five broad bands 
 are $3551$, $4686$, $6165$, $7481$, and $8931$ \AA{} \citep{fuku96,hogg01,smit02}. 
For details regarding the SDSS camera see \citet{gunn98}; for astrometric 
calibrations see \citet{pier03}. 
The survey has spectroscopy over 9380 sq. deg.; the spectroscopy is now 
complete over a large contiguous area of the Northern Galactic Cap, closing the gap 
that was present in previous data releases. 

In this work we employed spectroscopic and photometric data extracted from SDSS-DR7. 
The spectroscopic data were derived from the Main Galaxy Sample 
(MGS; \citet{mgs}) obtained from the \texttt{fits} files at the SDSS home 
page\footnote{http://www.sdss.org/dr7/products/spectra/getspectra.html}. 
k-corrections for this sample were calculated using the software 
\texttt{k-correct\_v4.2} of \citet{kcorrect}. 
The photometric data were derived from the photometric catalogue constructed by 
\citet{photo}\footnote{Available at http://casjobs.starlight.ufsc.br/casjobs}. 
These authors compute photometric redshift  and k-correction for the photometric data of the SDSS-DR7. 
The $rms$ of the photometric redshift is $\sigma_{phot} \sim$ 0.0227 and k-corrections were obtained 
through joint parametrisation of redshift and reference frame (at $z=0.1$) $(g-r)$ colour. 
For both data sets, k-corrected absolute magnitudes were calculated from Petrosian apparent magnitudes 
converted to the AB system.

\begin{figure*}
\begin{picture}(450,240)
\put(0,0){\psfig{file=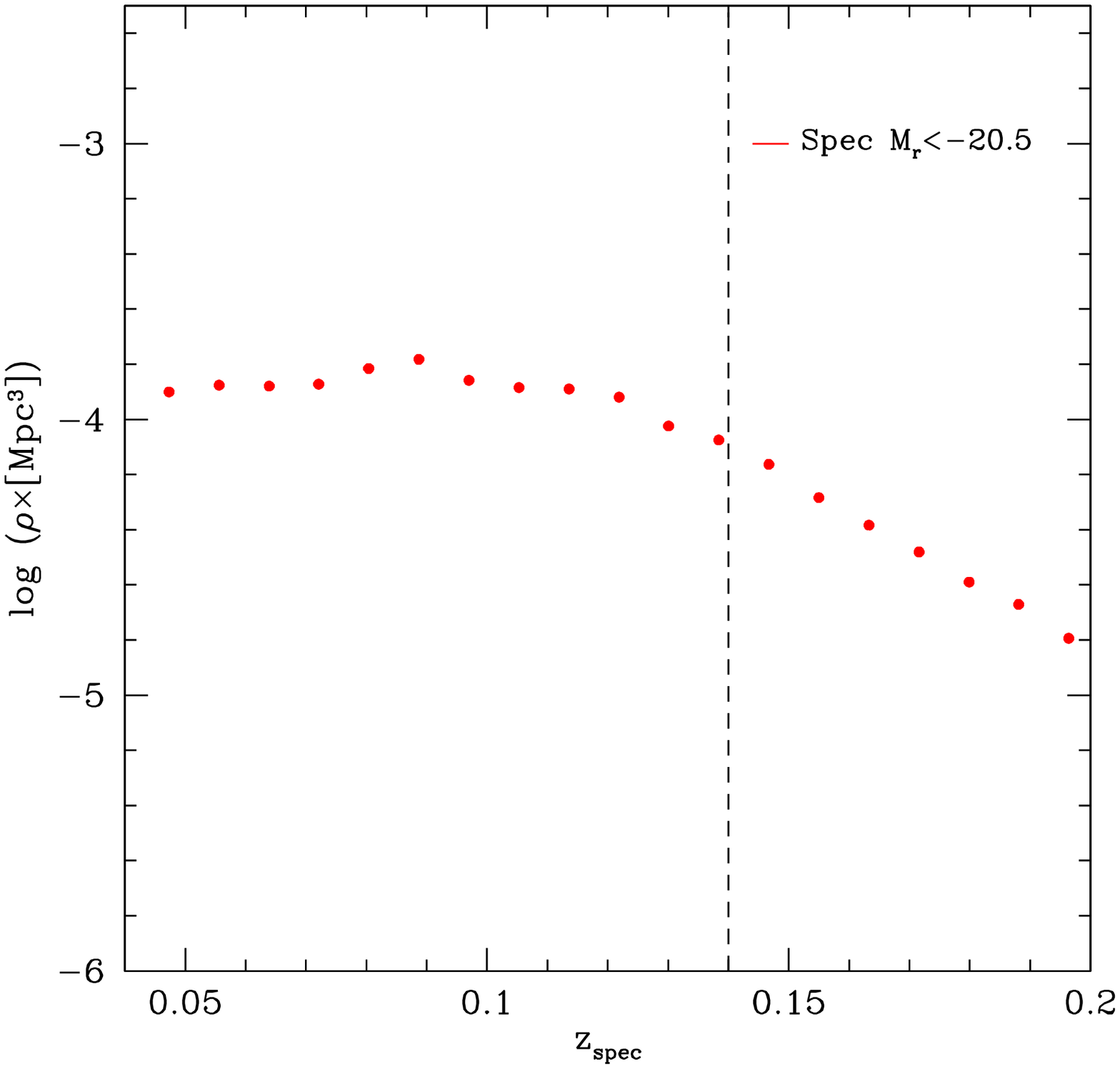,width=8cm}}
\put(240,0){\psfig{file=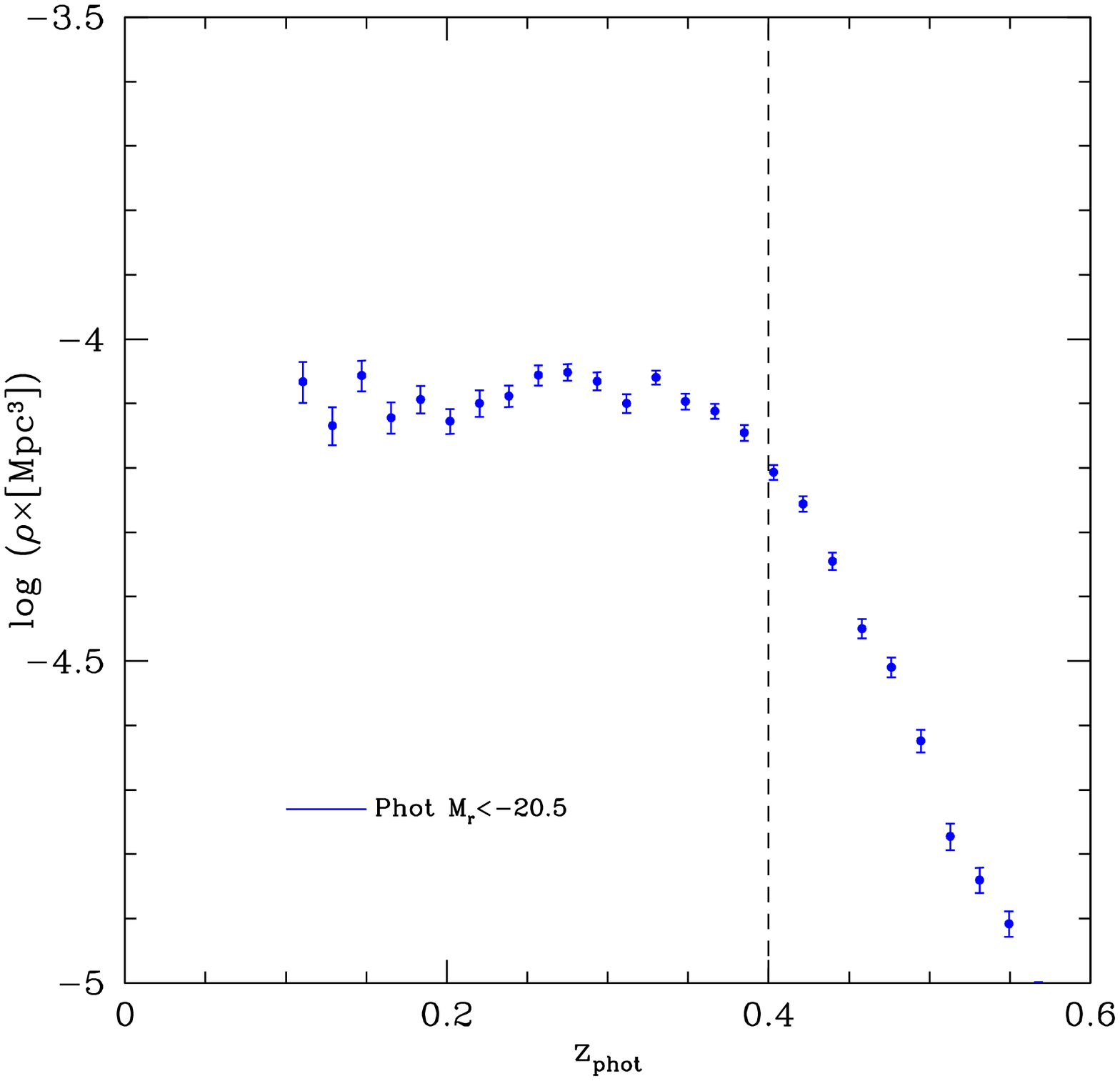,width=8cm}}
\end{picture}
\caption{Spatial number density of galaxies brighter than $ M_r =- 20.5 $. 
{\it Left:} Spectroscopic data with apparent magnitude $ r <17.77$.
{\it Right:} Random sample of $\sim0.001\%$ of the photometric sample with apparent magnitude $ r <21.5$.
Error bars represent $ 1 \sigma $ uncertainties   
calculated by bootstrap. The vertical line shows the limits chosen for the detection of the triple candidate systems.}
\label{f1}
\end{figure*}

In order to use these spectroscopic and photometric data, it is 
necessary to analyse the completeness and reliability of these samples. 
To explore this issue we have calculated the spatial number density of galaxies 
brighter than $M_r=-20.5$ for both spectroscopic and photometric data, taking into account the apparent 
magnitude limit $r=17.77$ and $r=21.5$, respectively. In Figure \ref{f1} it can be appreciated 
that the spatial number density of galaxy for, both, spectroscopic and photometric samples, shows a flat trend up to $z\sim 0.14$ and $z\sim 0.4$ respectively. 
The spatial number density of galaxies for the photometric data shown in Figure 1 corresponds to a random 
sample with $\sim 0.001\%$ of the complete photometric sample.

Taking into account these considerations we derived four galaxy data sets from SDSS-DR7:

The first sample, hereafter S1, comprises galaxies brighter than $r=17.77$ with spectroscopic redshift in the range  $0.01 \le \zspec<0.14$. This sample was employed for the detection of triple systems of galaxies candidates at low redshift.

In order to analyse the effect of photometric errors over the radial velocity differences $\Delta V$ for the photometric data, we have selected a sample that comprises galaxies brighter than $r=17.77$  with spectroscopic measurements in the redshift range $0.07\le \zspec \le 0.12$ for which we have obtained photometric redshifts. We will refer to this sample hereafter as the S2 sample.

We have also extracted a sample of galaxies brighter than $r=17.77$, that comprises 
galaxies with spectroscopic measurements in the redshift range $0.09\le \zspec \le 0.11$ and galaxies 
with photometric redshifts in the range $0.09-\sigma_{phot} \le \zphot \le 0.11+\sigma_{phot}$ (hereafter S3). 
This sample was employed in the analysis of completeness due to fiber collision, described in Section 4. 

Finally, for the detection of triplets of galaxies candidates at intermediate redshift we employed a sample derived » from the photometric catalogue of \citet{photo}. 
We have selected a sample that comprises galaxies brighter than $r=21.5$, 
in the redshift range $0.09\le z_{phot}\le 0.5$. We will refer to this sample as S4 hereafter.

Table 1 summarises the main characteristics of the different
samples used in this work.

\begin{table*}
\begin{minipage}{175mm}
\caption{Description of the samples used in this work.}
  \begin{center}\begin{tabular}{@{}cl@{}cl@{}}
  \hline
  \hline
sample name &  Description  & number of objects\\
 \hline
S1 &  MGS spectroscopic data.  & 684939 \\
    &  Used for the identification of triplets of galaxies candidates at low redshift ($0.01\le \zspec<0.14$)\\
    \hline
S2 & MGS with photometric redshift ($0.07 \le \zspec \le 0.12$). & 248210 \\
    & Used in section 4.3\\
\hline
S3 & MGS and photometric data. & 80851 and 860021 \\
    & Used in the fiber collision analysis\\
\hline
S4 &  Photometric catalogue with redshifts and k-correction from \citet{photo}  & 66390174 \\
    &     Used for the identification of triplets of galaxies candidates at intermediate redshift ($0.09\le \zphot \le 0.5$)\\
\hline
\hline
\label{t1}
\end{tabular}
\end{center}
\end{minipage}
\end{table*}


\section{Building the catalogue}

In this section we present the algorithm adopted to identify galaxy triplet candidates 
from the spectroscopic and photometric volume limited samples of SDSS-DR7. The algorithm
searches for triplets by selecting galaxies within a given projected distance ($r_p$) 
and radial velocity difference ($\Delta V$) of another galaxy.

 \citet{Lambas03} identified pairs of 
galaxies using $r_p=100\kpc$ and $\Delta V=350\kms$. \citet{manuel} identified 
galaxy groups in the SDSS third data release with a mean redshift of 0.1 and a 
median velocity dispersion of $230\kms$. These authors employed a fiends-of-friends algorithm with a 
transverse linking length corresponding to an overdensity of 80 and a line-of-sight 
linking length of $\Delta V=200\kms$. For the detection of triple galaxy systems, 
we have considered these restrictions in the selection of $r_p$ and $\Delta V$ constrains.


\subsection{Data Pixelization}

For the purpose of a suitable selection of the systems, as well as to efficiently reduce the computing time, 
we have used a pixelization of SDSS data. Indeed, data pixelization is an important tool 
in managing large amounts of information such as that contained in the SDSS photometric catalogue.

In order to perform a pixelization of SDSS data, we have employed routines of the Hierarchical Equal Area Pixelization isoLatitude software
\citep[HEALPIX,][]{healpix}. HEALPIX was developed for data processing and analysis of the observations of the
cosmic microwave background (CMB). This pixelization procedure consists in an equal area 
isolatitud partition of the sphere, which results into a versatile structure for data analysis. 
The resolution base comprises 12 pixels formed by three rings around the poles and the equator. 
For higher resolutions, each pixel is subdivided into four equal-area pixels of smaller size. 
The grid resolution is set by the $N_{side}$ parameter, which defines the number of 
divisions along the side of a pixel, required to achieve the high-resolution partition desired. 
All pixel centres are equidistant in azimuth for each ring and are placed on rings of constant latitude. 
All rings located in the equatorial zone are divided in the same number of 
pixels, $N_{eq}=4N_{side}$. The rings located in the polar cap region contain a varying 
number of pixels that increase by one pixel within each quadrant, with increasing distance 
from the poles. 
The resulting map contains $N_{pix}=12N_{side}^2$ pixels of equal area $A=\pi/N_{side}^2$.

HEALPIX has a library of computational algorithms and visualisation software that allows
fast scientific applications on discretized maps created from large amounts of data as is the SDSS survey.

We have employed HEALPIX routines to find the index of all pixels within a radial 
angular distance from a defined center. The routines used were: $ ang2pix $, which 
generates a linking list for galaxies in each pixel; $ang2vec$, that  converts angular 
to Cartesian coordinates, and, finally, $query\_disc$, that identifies the central galaxy 
pixel and the adjacent pixels inside the search radius, previously defined. 
The resolution ($ N_ {side} $) was set to 512 at redshift smaller than 0.14 (S1, S2 and S3 samples) 
and 2048 at redshift greater than 0.14 (S4 sample). 
The choice of different values of $ N_ {side} $ is due to the density difference between the
lower and intermediate redshift samples. 

For each selected center we define a search radius that covers approximately $600\kpc$ at 
the smallest redshift of each sample, ensuring the search in every pixel adjacent to the center. 
The adopted search radii were: $ 0.88\deg $ for S1, S2 and S3 samples and $0.068\deg$ for the S4 sample.



\begin{figure*}
\leavevmode \epsfysize=10cm \epsfbox{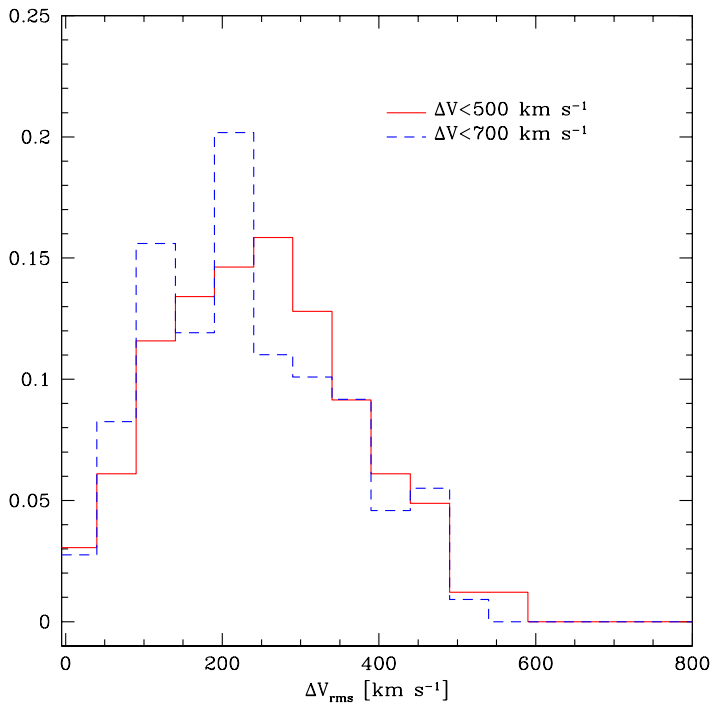}
\caption{Distribution of $\Delta V_{rms}$ for triplet candidates identified in the
spectroscopic sample S1. 
Red (continuous) line:  $r_p<100 \kpc$ and $\Delta V<500 \kms$. Blue (dashed) line: 
systems with $r_p<200\kpc$ and $ \Delta V<700\kms$.} 
\label{f3}
\end{figure*}

\subsection{The algorithm}

In this section we describe the successive steps of the algorithm developed for 
the identification of triplet of galaxies candidates. 

\textbf{Step 1:} The algorithm begins with a selection of central galaxy candidates. The pixel 
of each candidate is identified. 
Then a search for galaxies belonging to adjacent pixels, enclosed within the 
search radius, is performed. In the first run, any galaxy is considered a central galaxy candidate. 

\textbf{Step 2:} To identify triplet systems, we have considered as neighbours all those 
galaxies that satisfy the constraints in projected distance and radial velocity difference, ie.  
$r_p<r_{pmax}$ and $\Delta V<\Delta V_{max}$. 

\textbf{Step 3:} We consider a triplet candidate when there are only three galaxies within the 
constraints in Step 2.

\textbf{Step 4:} Then, we assign the brightest galaxy to the center of each system. 
This generate two types of multiple identifications:
\begin {itemize}
\item The center is assigned to more than one system independently identified.
This is because the brightest galaxy is a neighbour galaxy in more than one system.
\item A system will be identified more than once if different centres satisfy the neighbourhood
condition, ie have $r_p< r_{pmax}$ and $ \Delta V< \Delta V_{max}$.
\end {itemize}
In order to avoid double identifications, repeated centres were removed from the sample 
and close galaxy pair centres were considered as a single system by taking the 
brightest galaxy as a central galaxy. This procedure generates a new sample of central galaxies. 
With these centres as  input sample, the algorithm returns to Step 1.
 
\textbf{Step 5:} It may happen that the new central galaxies may be located at a distance 
greater than $r_{pmax}$ or have a radial velocity difference greater than  $\Delta V_{max}$ 
with respect to one of the selected neighbours. To retrieve these systems, we consider a second 
set of constraints, relaxing restrictions over velocity difference and projected distance, 
and then returning to Step 2.



\section{Implementation of the algorithm at low redshift $(0.01\le\zspec<0.14)$}

The aim of this work is the identification of triplets of luminous galaxies, 
defined as systems close in projected distance and radial velocities. 
Building of a catalogue of triplets with spectroscopic measurements 
is less subject to contamination than using photometric redshifts given the small 
uncertainties. So, in principle, using spectroscopic data results into a more 
reliable identification of these objects.

\subsection{Spectroscopic data}

In order to identify triplets of galaxies candidates in the S1 sample, 
we have run the algorithm considering the following restrictions for the selection 
of neighbouring galaxies: $ r_p <100\kpc $ and $ \Delta V <500\kms$.

Following the steps described in the previous section, we have assigned the center to 
the brightest galaxy of the system. After this recentering,  we have considered a second 
group of constrains $ r_p <200\kpc $ and $ \Delta V <700\kms$, in order to retrieve the 
systems lost by  the procedure. 

Therefore, a total of $273$ triple system candidates were obtained from the S1 sample, $164$ of 
them with the constraints $ r_p <100\kpc $ and $ \Delta V <500\kms $ and $109$  with 
$r_p <200\kpc $ and $ \Delta V<700\kms $. 

Figure \ref{f3} shows the distribution of $\Delta V_ {rms}$, where $\Delta V_ {rms}$ is the root mean 
square radial velocity difference for the systems. Red lines correspond to the 
triplet candidates obtained whit $ r_p <100\kpc $ and $ \Delta V <500\kms $ and in blue 
lines we present the systems obtained with $ r_p <200\kpc $ and $ \Delta V <700\kms$. 
Both distributions present a similar behaviour.

 \begin{figure*}
\begin{picture}(450,240)
\put(0,0){\psfig{file=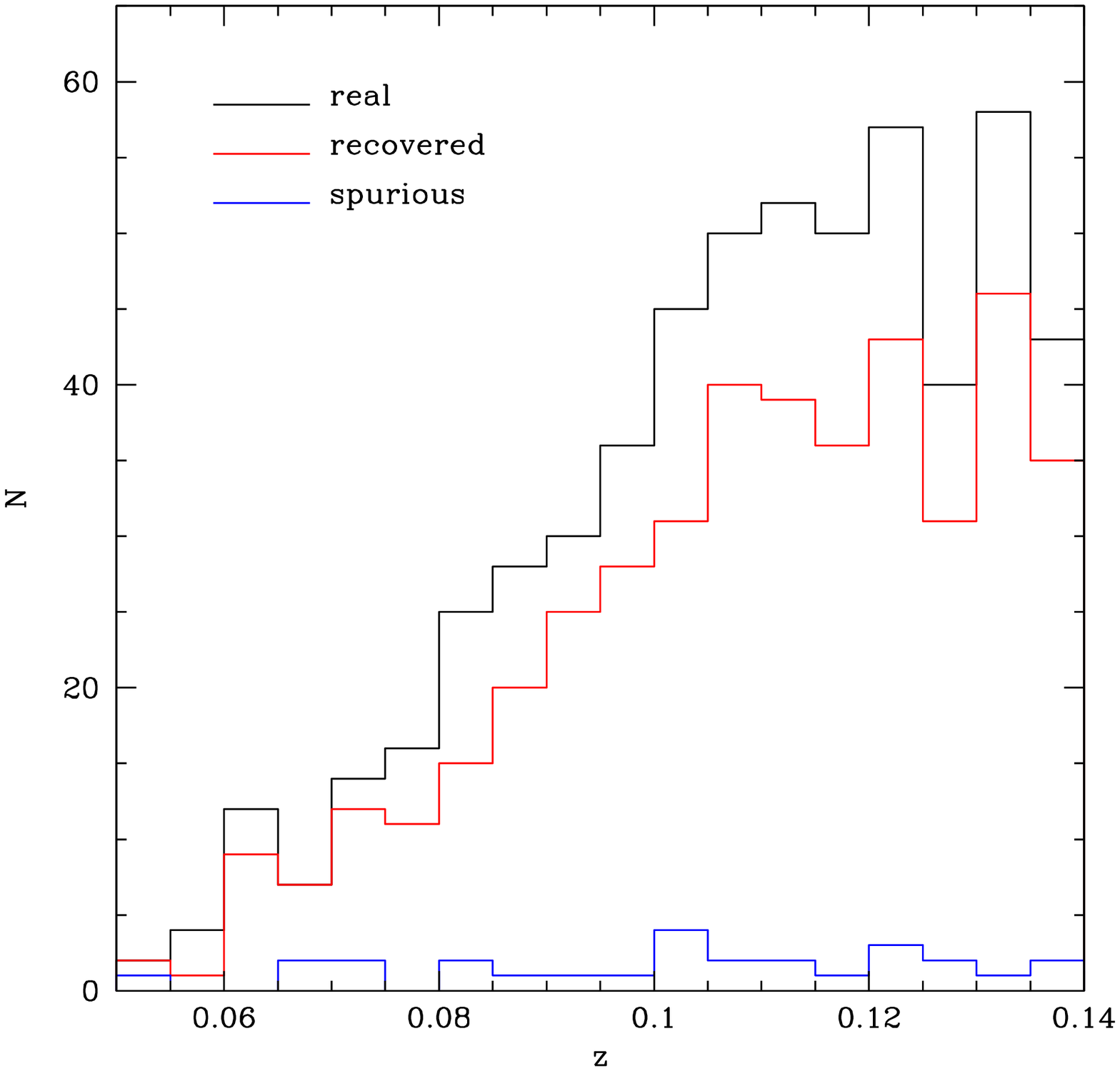,width=8cm}}
\put(240,0){\psfig{file=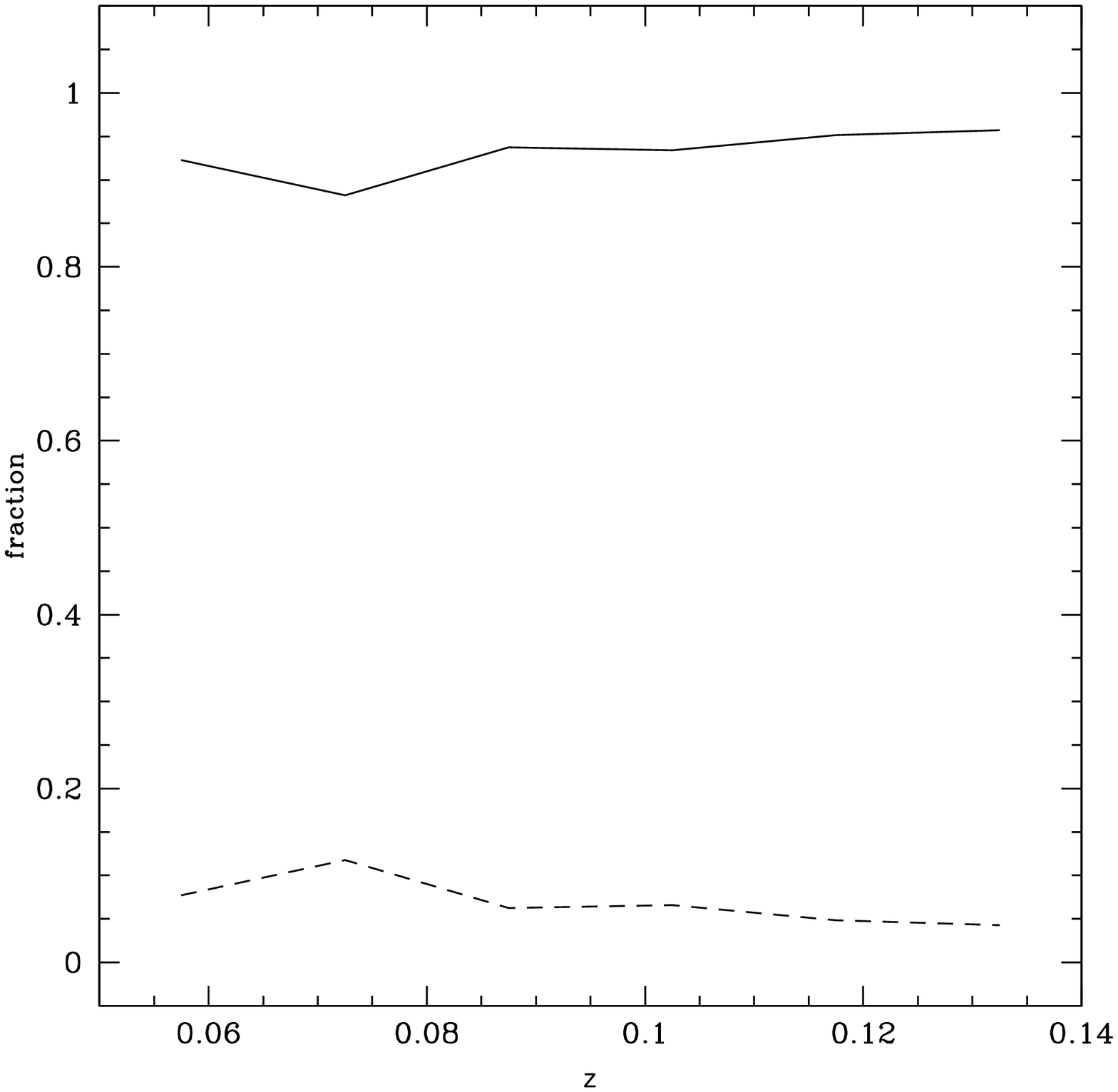,width=8cm}}
\end{picture}
\caption{Completeness and contamination analysis. Left panel: Real, recovered and spurious systems  
redshift distributions. Right panel: Completeness (solid line) and contamination (dashed line) 
as function of redshift.} 
\label{f4}
\end{figure*}

\subsection{Completeness and contamination tests}

In order to test the algorithms, we performed a completeness and contamination test 
employing a SDSS mock catalogue derived from the semi-analytic models of \citet{croton}, obtained
from the Millennium Simulation \citep{springel} outputs. The spatial resolution of this 
simulation is suitable for the implementation of our algorithm. The luminosity function of the 
semi-analytic model is consistent with observations of galaxies at low redshift. 
The mock catalogue comprise 514914 galaxies brighter than $r=17.77$, where $r$ correspond to 
the $r-band$ magnitude of the SDSS. This mock catalogue contains: right ascension, declination, 
apparent magnitude in $r-band$, halo mass, redshift and peculiar velocity. 
 
For the completeness and contamination test we have employed a 
volume limited mock catalogue considering galaxies brighter than $M_r=-20.5$ in 
the redshift range $0.01\le z<0.14$.

The completeness is defined as ${N_{rec}}/{N_{tot}}$, where $N_{rec}$ is the number of triplet candidates  
identified with our algorithm that are matched to the 
systems found in real space (recovered triplets), and $N_{tot}$ is the number of systems selected with real 
distances (real triplets) \footnote{$r_p=r \times cos(\theta)\times d\theta \times d\Phi$, where $r$, 
$\theta$ and $\Phi$ are the spherical coordinates with origin in the triplets center.} within the same halo. 
We find that the completeness achieved by our algorithm is $\sim 80\%$. 

\begin{figure*}
\begin{center}
\leavevmode \epsfysize=10cm \epsfbox{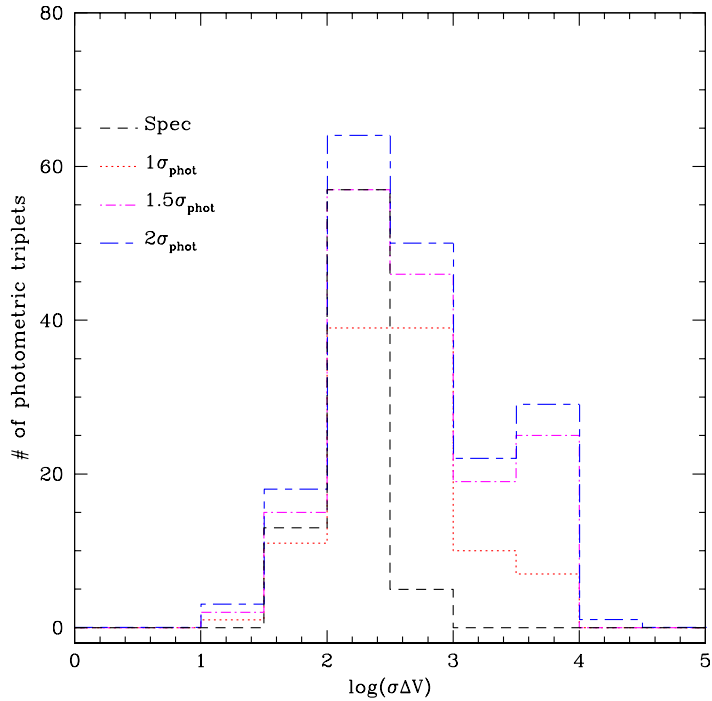}
\end{center}
\caption{$\sigma_{\Delta V}$ spectroscopic distribution for photometric triplet candidates using S2 sample.
Black (short dashed line): spectroscopic systems. Red (dotted), magenta 
(dot-short dashed line) and blue (short dashed-long dashed line): photometric systems obtained
considering  $1\sigma_{phot}*c$, $1.5\sigma_{phot}*c$ and $2\sigma_{phot}*c$ velocity intervals, respectively.} 
\label{f5}
\end{figure*}

The contamination is defined as ${N_{sp}}/{N_{det}}$, where $N_{det}$ is the number 
of triplet candidates identified with $r_p$ and $\Delta V$. $N_{sp}$ is number of systems detected whose 
member galaxies do not belong to the same halo (spurious triplets). The contamination for our triplet 
sample is very small, $\sim 5\%$.

Figure \ref{f4} shows the redshift distribution for the real, 
recovered and spurious triplet of galaxies, as well as the dependence of
the completeness and contamination rates with redshift.

%


\subsection{Analysis of photometric data in the spectroscopic redshift range}

The major source of contamination in the detection of the photometric systems is due to 
the error in $\zphot$, since the uncertainties in photometric redshifts affects the 
choice of $\Delta V$. To evaluate this point, we employed the S2 sample. 
We have run the algorithm for different values of $\Delta V$, corresponding 
to $1\sigma_{phot}*c$, $1.5\sigma_{phot}*c$ and $2\sigma_{phot}*c$, where $\sigma_{phot}$ 
is the mean photometric redshift error ($\sigma_{phot} \sim 0.0227$, as previously mentioned in Section 2) 
and $c$ is the speed of light. 
We have then examined the distribution of the
values of the $rms$ spectroscopic velocity differences
$\sigma_{\Delta V}$, considering the spectroscopic redshifts of the triplet candidate members identified
photometrically.

For the S2 sample we identified 80 triplet candidates that satisfy the constraints described 
en section 4.1. Using photometric data we recover 71 of these systems considering 
$1\sigma_{phot}$ and $1.5\sigma_{phot}$ for $\Delta V$. 
For $2\sigma_{phot}$ we recovered all the spectroscopic triplets.

In figure \ref{f5} we show the $\sigma_{\Delta V}$ distribution of triplet candidates identified 
in S2 sample. In different lines and colours we show the systems identified with 
$1\sigma_{phot}$ in red, $1.5\sigma_{phot}$ in magenta, and $2\sigma_{phot}$ in blue line. 
From the figure it can be appreciated that we have a large level of contamination when using  
$1.5\sigma_{phot}$ and $2\sigma_{phot}$. Therefore, we conclude that the $1\sigma_{phot}$ interval provides a 
suitable compromise between high completeness and low contamination.


\subsection{Incompleteness due to fiber collisions}

The SDSS spectrograph uses fibers manually connected to plates in the telescope's focal plane. 
These fibers are  mapped  through a mosaic algorithm \citep{blanton2003} that optimises the 
observation of large-scale structures. Two fibers can not be placed closer than 55''  
\citep{mgs}, so 
for two objects with the same priority (such as two MGS galaxies) and whose centres are closer than 55'', 
the algorithm selects at random the galaxy which will be observed spectroscopically. 
There are regions where the plates overlap (about $ 30\%$ of the mosaic regions), 
in which both objects may be observed.

Due to fiber collision the spectroscopic sample is affected by incompleteness.
The magnitude limit of spectroscopic objects is $r=17.77$, but not all galaxies brighter than this 
limit were observed. This issue becomes more important when 
analysing compact objects.

\begin{figure*}
\leavevmode \epsfysize=9cm \epsfbox{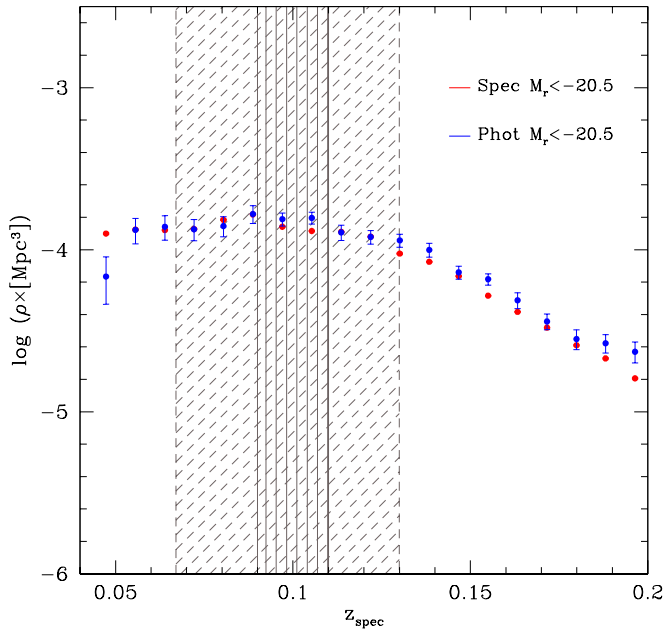}
\caption{Spatial number density of galaxies as a function redshift for the MGS (red dots) and for a random sample 
with $0.001\%$ of the photometric sample from \citet{photo} (blue dots) both with $r<17.77$. 
The error bars represent $1\sigma$ uncertainties 
calculated using bootstrap. For both samples the galaxy density for $\zspec<0.14$ is 
nearly flat. 
The small and large regions correspond to redshift intervals employed in the analysis 
of fiber collision (S3 sample).}
\label{f6}
\end{figure*}

\citet{zehavi02} found that approximately $6\% $ of the target objects are not assigned to a fiber 
due to the 55'' restriction. \citet{alonso06} have identified pairs of galaxies in the MGS of SDSS second 
data release and analysed galaxy pairs loss due to fiber collision. These authors found 
that approximately $5\%$ of the pairs were not recovered due to this effect. Recently \citet{mvcd11} found that $\sim 40\%$ of galaxies with $r<17.77$ have spectroscopic measurements.


In figure \ref{f6} we show the spatial galaxy density as a function of redshift for the MGS 
(red dots) and for a random sample with $0.001\%$ of the photometric data from \citet{photo} 
(blue dots) both with $r<17.77$. It can be appreciated a nearly flat galaxy density for $\zspec<0.14$. 
In order to evaluate the reliability of the identification procedure adopted
in this work, we used the S3 sample considering spectroscopic and photometric triplets detected in the 
redshift range $0.09\le \zspec \le 0.11$ (figure \ref{f6} small region) and 
$0.09-\sigma_{phot}\le \zphot \le 0.11+\sigma_{phot}$ (figure \ref{f6} large region) respectively. 
For a discussion about the uncertainties in the photometric redshift see subsection 4.3.

Using S3 sample we have obtained $64$ triplet candidates with spectroscopic measurements.  
About $\sim 90\%$ of the systems had been recovered using photometric data. 
Nevertheless, there are $39$ triplets with 1 or 2 members without spectroscopic 
information due to fiber collision. 
Figure \ref{f8} shows two examples of photometric triplet candidates where 
one or two galaxies do not have spectroscopic redshift determination. 

\begin{figure*}
\begin{picture}(450,240)
\put(0,0){\psfig{file=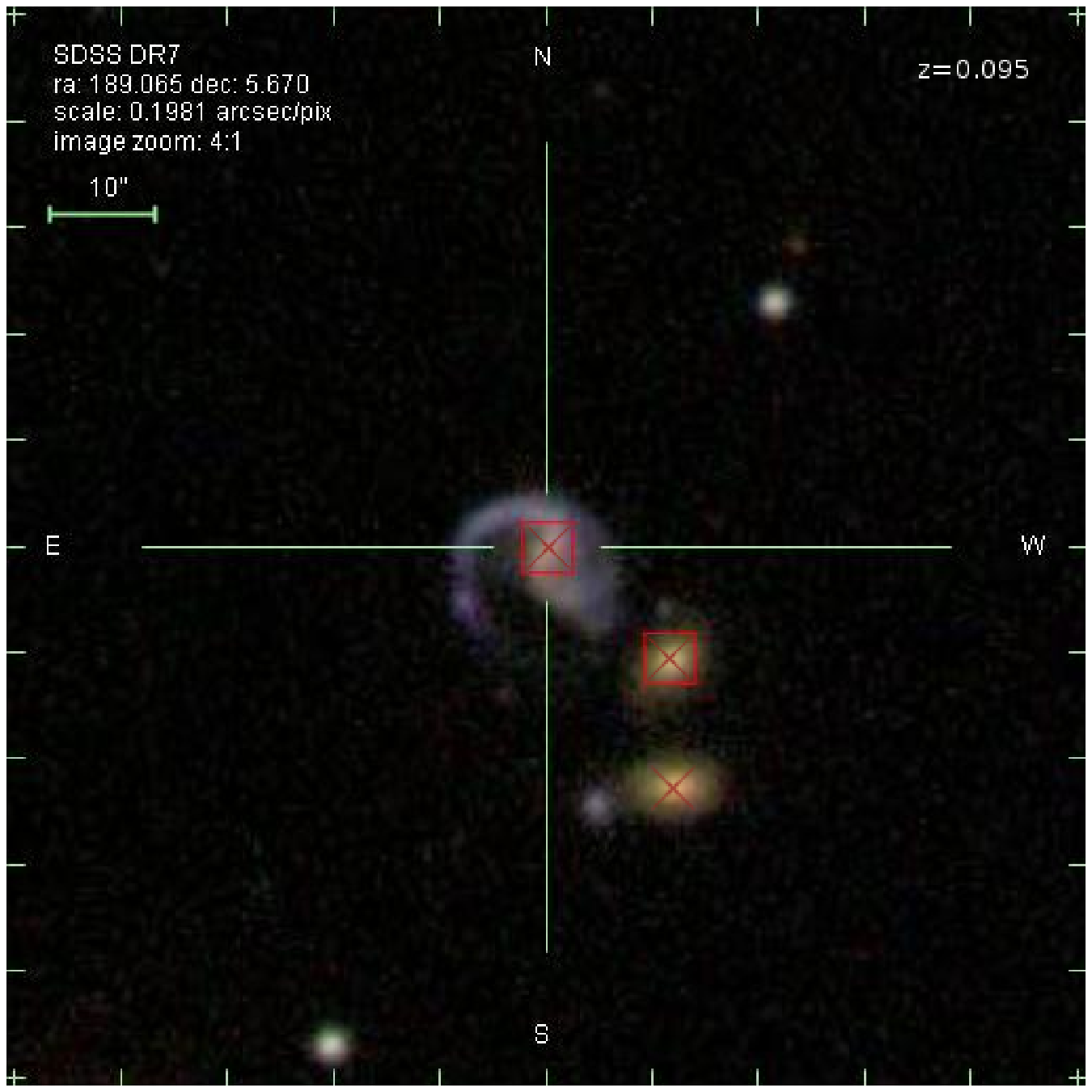,width=8cm}}
\put(240,0){\psfig{file=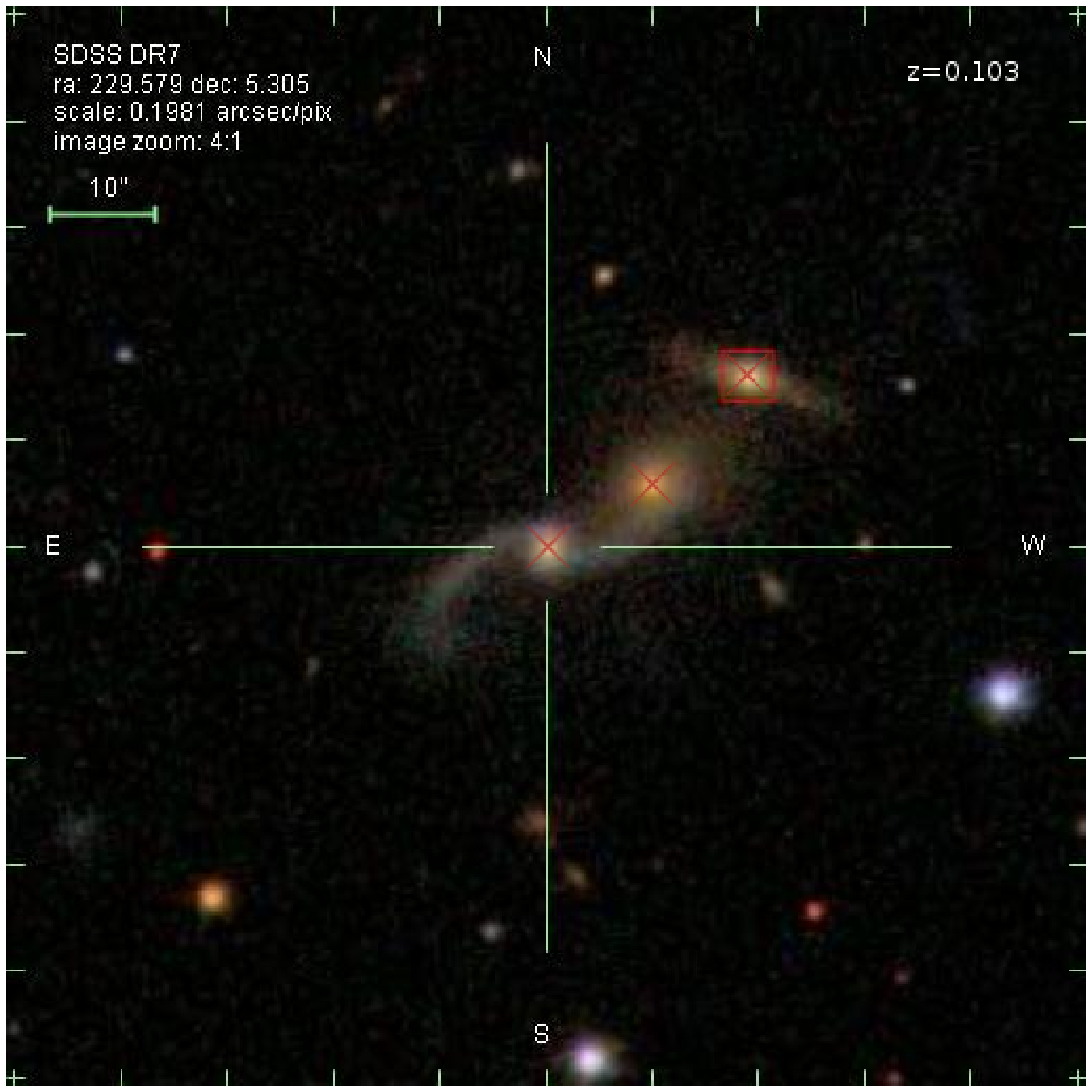,width=8cm}}
\end{picture}
\caption{Examples of triplet candidates with photometric measurements obtained from S3 sample. 
Galaxies with redshift determination are marked with a box, and galaxies members are 
indicate with crosses.}
\label{f8}
\end{figure*}
We have taken into account these considerations and we have included these objects 
as triplet of galaxies candidates at low redshift.


\section{Implementation of the algorithm at intermediate redshift ($0.14\le \zphot\le 0.4$)}

From the discussion in Sections 4.3 and 4.4 we can see that photometric redshifts 
are suitable to identify triplet of bright galaxies candidates. The right panel of figure \ref{f1} 
show the galaxy density distribution as a function of redshift for the photometric 
S4 sample. It can be noticed the flat behaviour of the galaxy density up to $\zphot <0.4$. 

The uncertainty of the photometric redshifts became larger at low redshift \citep{photo}. 
Taking into account these issues we assume a range $0.14\le\zphot\le 0.4$ of 
reliability for the determination of photometric triplet candidates. 

We run the algorithm for S4 sample considering the restrictions in $\Delta V$ according 
to the discussion of section 4.3.

Following the steps described in section 3.2, a total of $5628$ triplet system candidates were obtained in the range 
$0.14\le\zphot\le 0.4$, $3771$ of them with the constraints $ r_p <100\kpc $ and $ \Delta V <6810\kms$ ($1\sigma_{phot}\ast c$) 
and $1857$  with the second set of constrains $r_p <200\kpc $ and $ \Delta V<9310\kms$ 
(corresponding to $\sim 1.37\sigma_{phot}*c$).

\begin{figure*}
\leavevmode \epsfysize=10cm \epsfbox{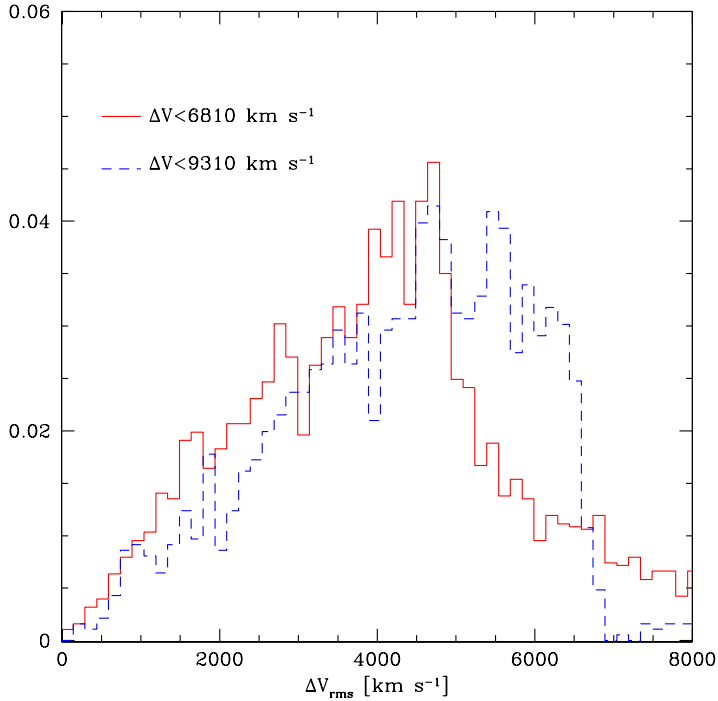}
\caption{Distribution of $\Delta V_{rms}$ for the systems identified in S4 sample. 
Red line:  $r_p<100 \kpc$ and $\Delta V<6810\kms$. Blue dashed line 
systems with $r_p<200\kpc$ and $ \Delta V<9310\kms$.} 
\label{f9}
\end{figure*}

Figure \ref{f9} shows the distribution of the values of $\Delta V_ {rms}$ for the systems identified 
in the S4 sample, with the higher $ \Delta V$ and $r_p$ constraints (red line), as well as this distribution
for systems identified relaxing these constraints (blue dashed line). 
From this figure it can be appreciated that both distributions show a similar behaviour.



\begin{figure*}
\begin{picture}(450,240)
\put(0,0){\psfig{file=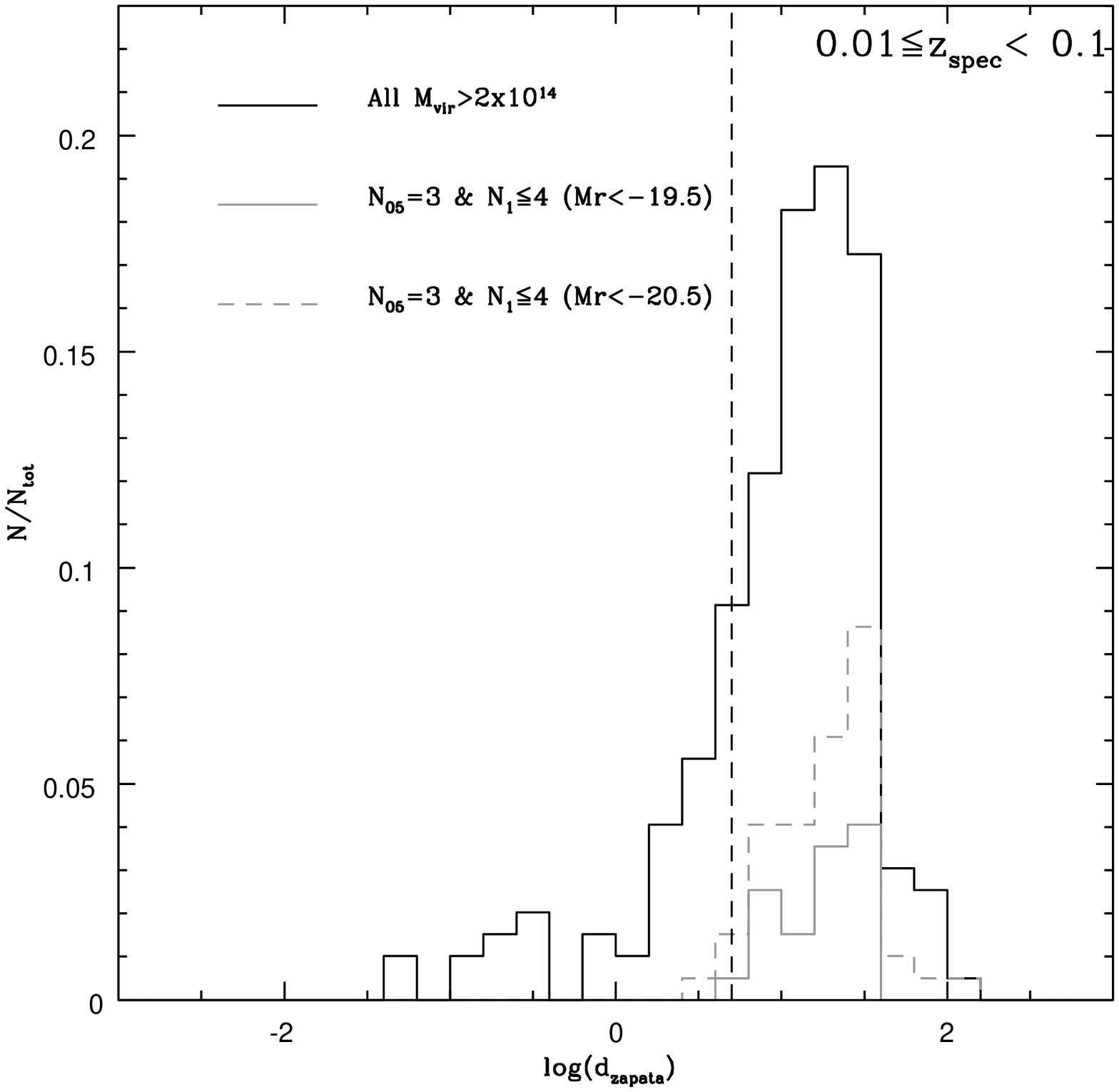,width=8cm}}
\put(240,0){\psfig{file=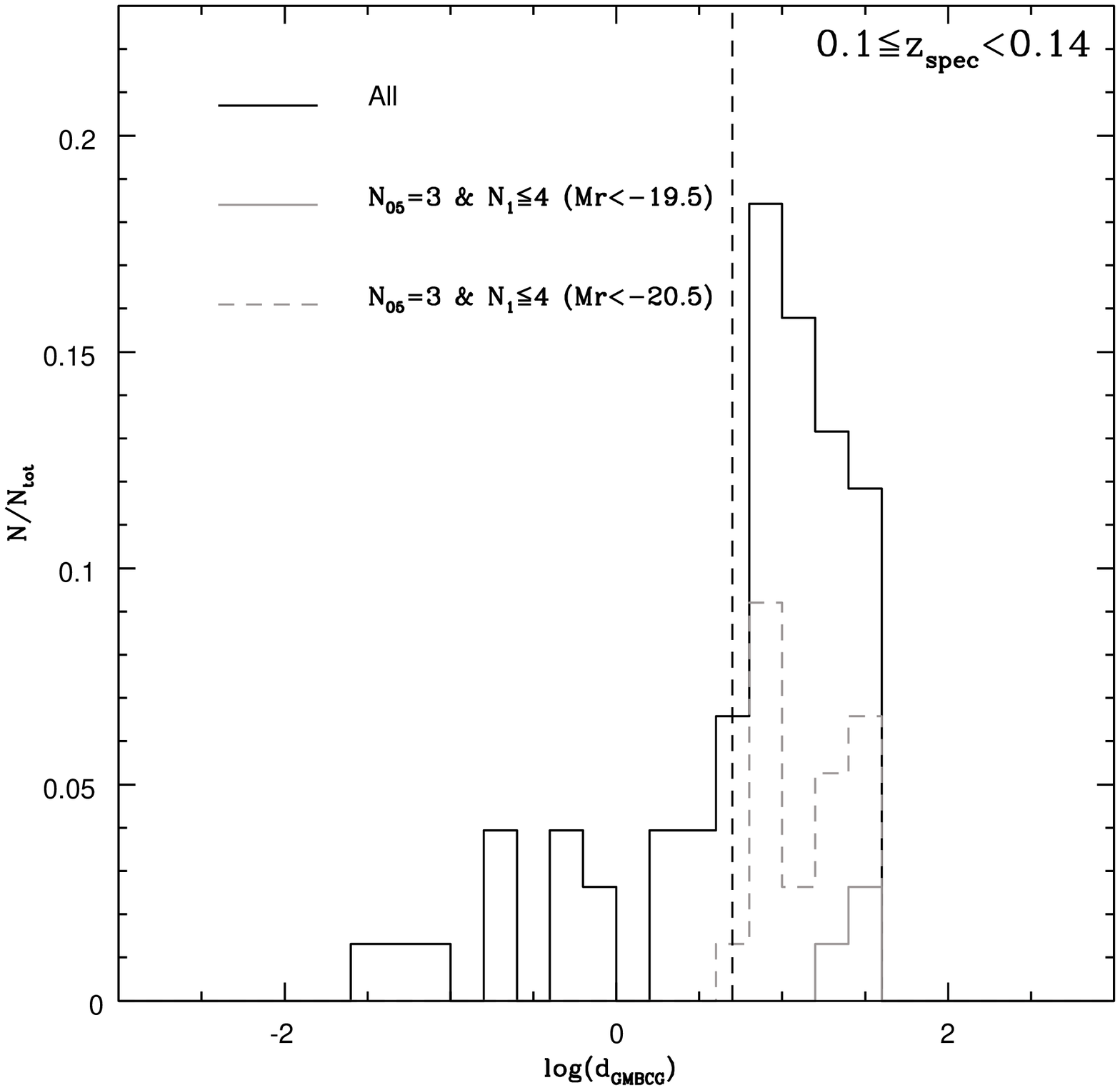,width=8cm}}
\end{picture}
\caption{Distribution of the distance to the closest neighbour cluster for spectroscopic triplet candidates. Left panel: Distance to \citet{zapata09} for systems with $0.01\le \zspec<0.1$. Right panel: Distance to GMBCG clusters for systems in the range with $0.1\le \zspec<0.14$. Solid black line show the distribution of all the systems in that redshift range, in grey dashed line the distribution of the triplet candidates with  $N05=3$ and $N1\le 4$ estimated using galaxies brighter than $M_r=-20.5$ and in solid grey line the systems with $N05=3$ and $N1\le 4$ estimated using galaxies brighter than $M_r=-19.5$. Vertical black dashed line show the $dc=5\mpc$ limit selected as part of the isolation criteria. }
\label{f8}
\end{figure*}

\section{Isolation criteria}

In the previous sections we have identified galaxy triplets candidates, regardless the relative 
spatial isolation of these systems. In order to build a catalogue of physical triplet systems, it is necessary to define isolation critera that ensures that 
the dynamics of these systems is not dominated by larger virialized structures where these 
systems could be immersed.

\begin{figure*}
\includegraphics[width=55mm,height=55mm]{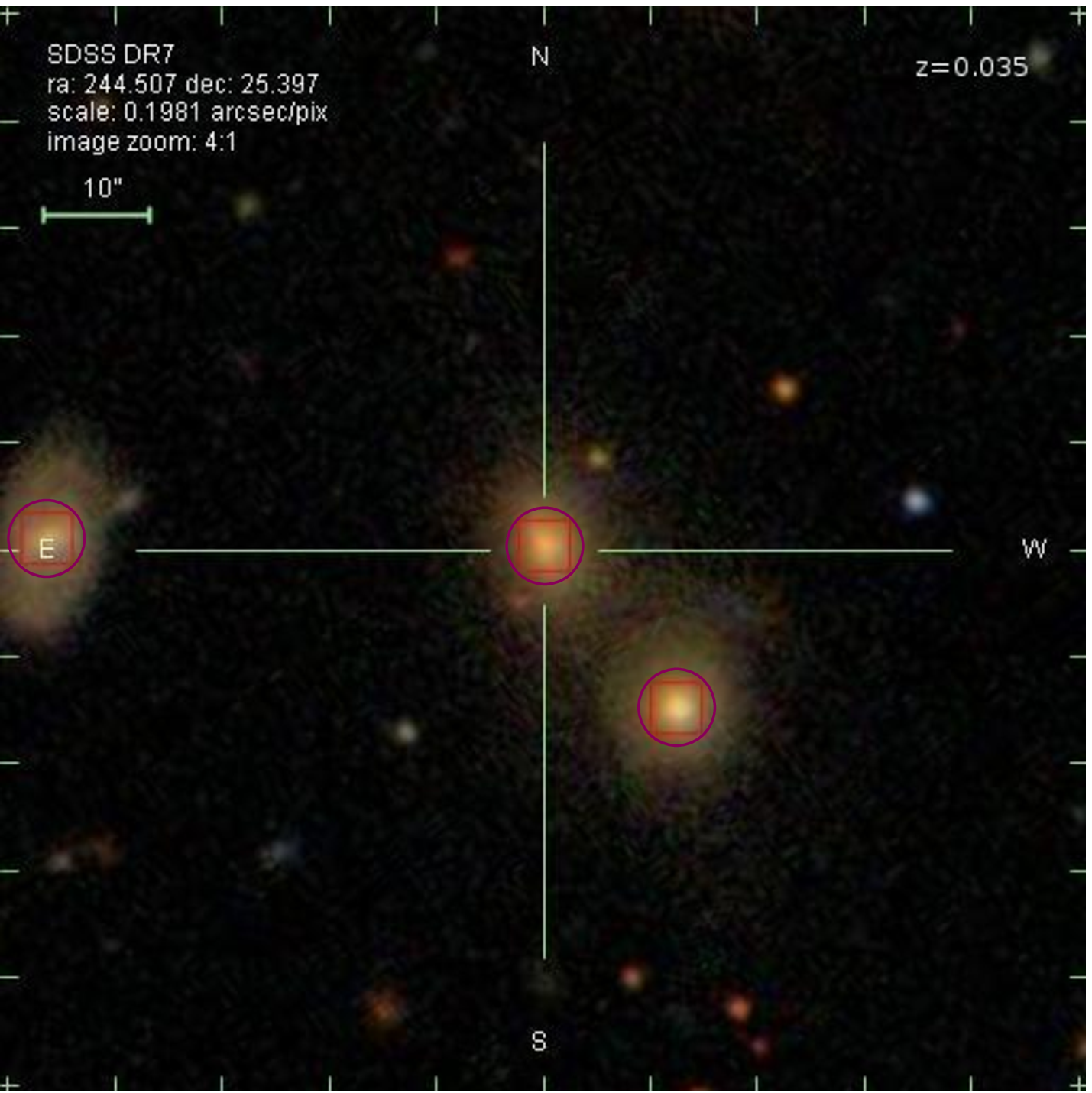} 
\includegraphics[width=55mm,height=55mm]{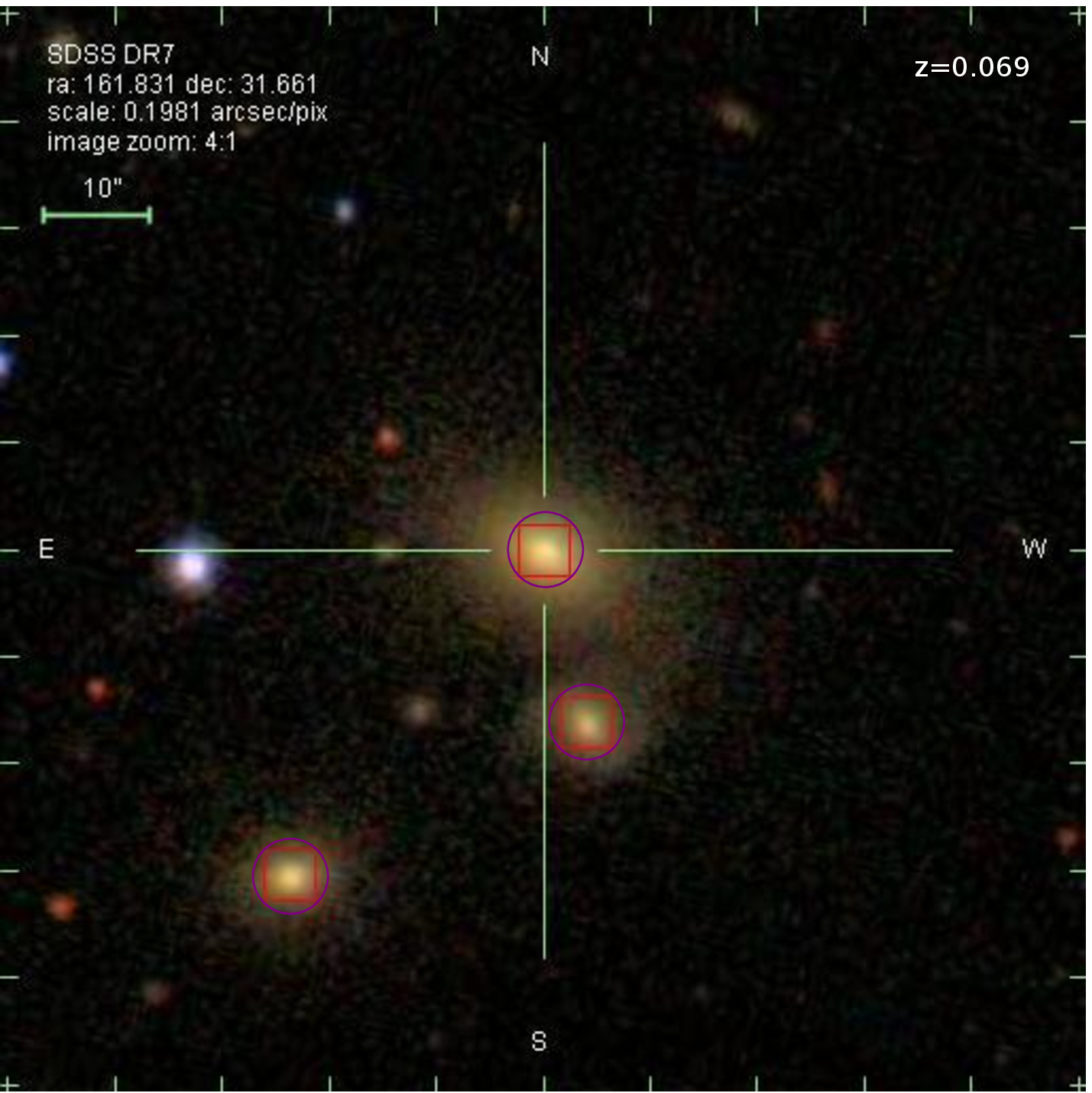}
\includegraphics[width=55mm,height=55mm]{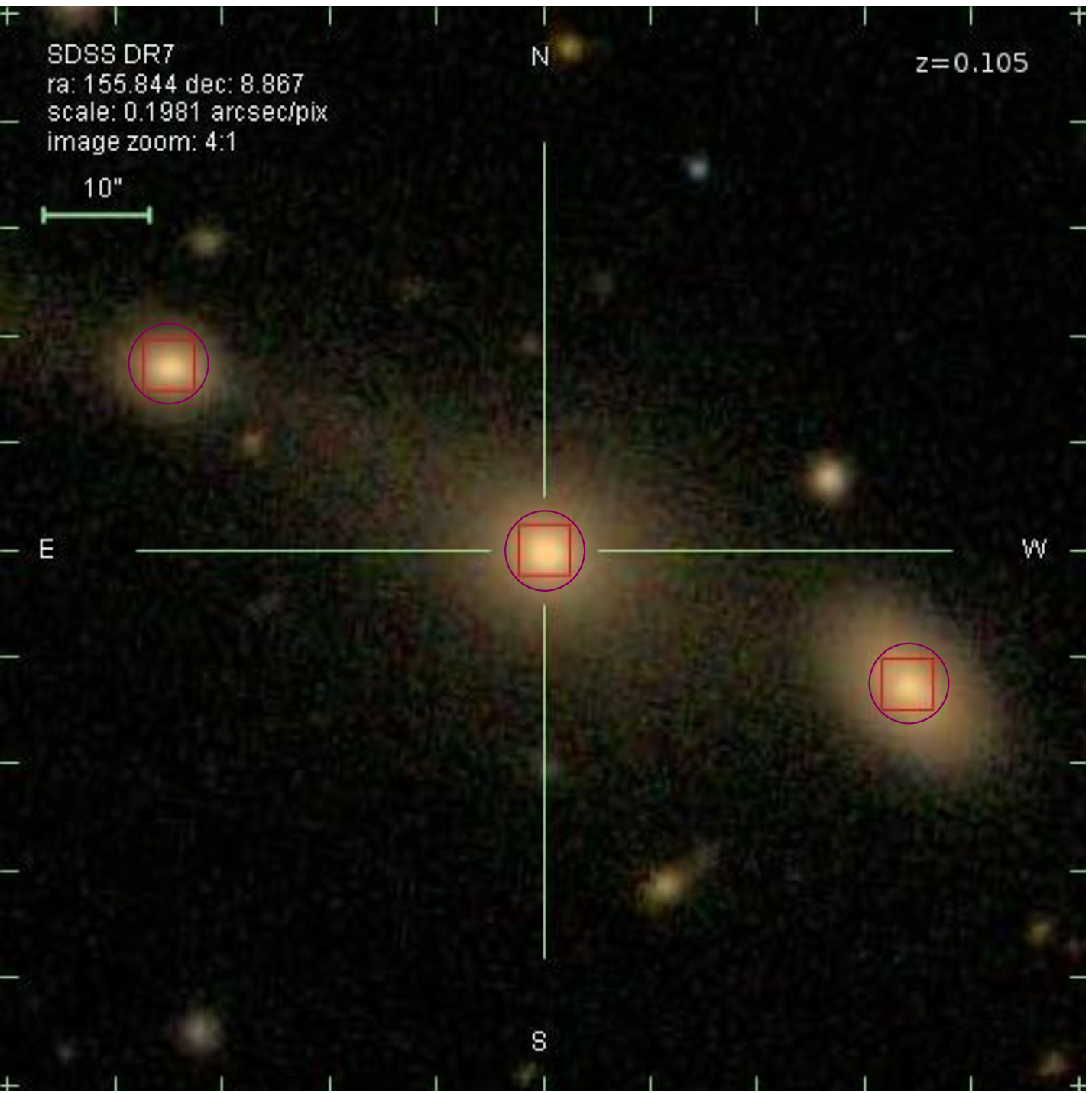} 
\includegraphics[width=55mm,height=55mm]{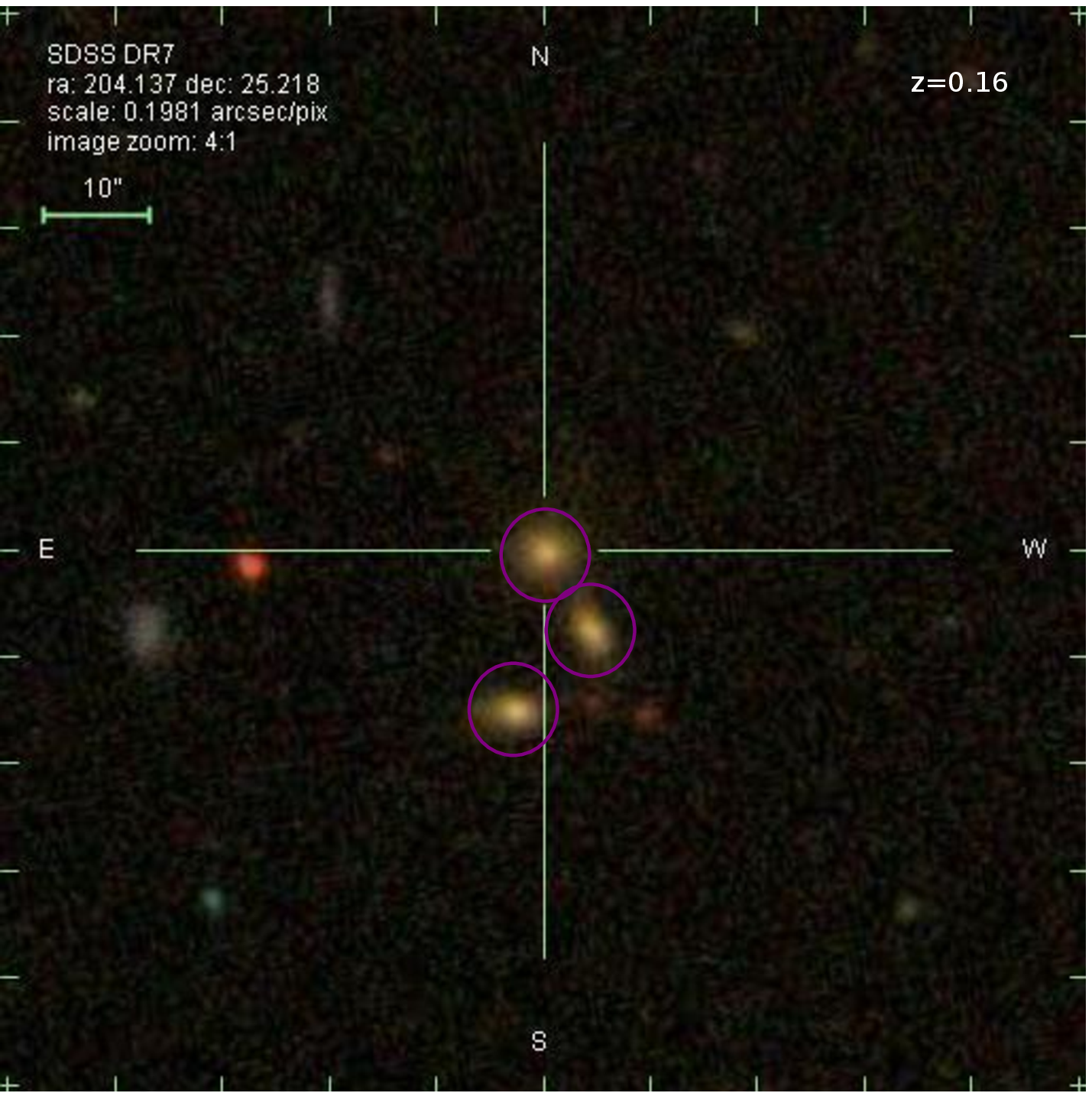} 
\includegraphics[width=55mm,height=55mm]{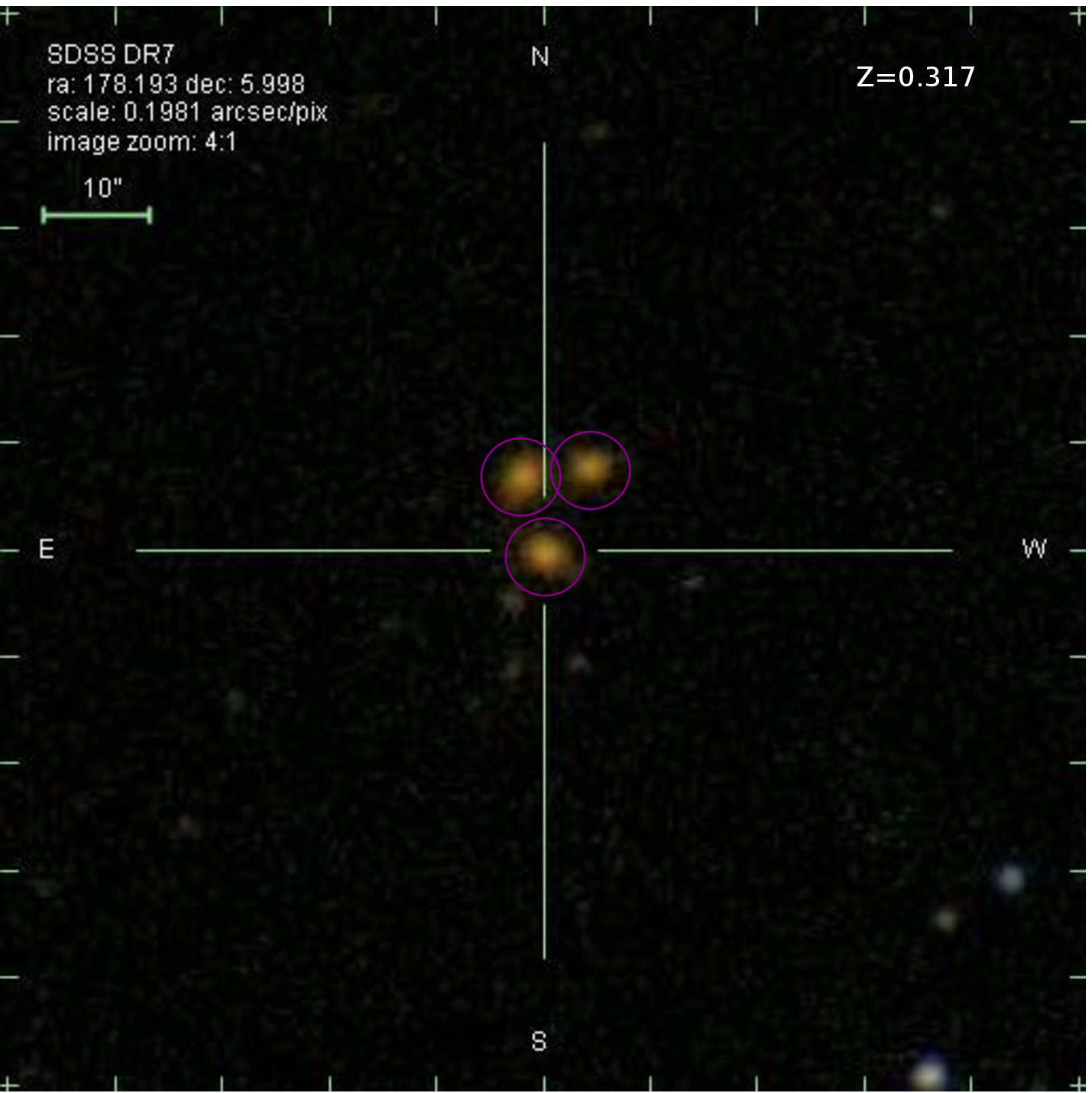}
\includegraphics[width=55mm,height=55mm]{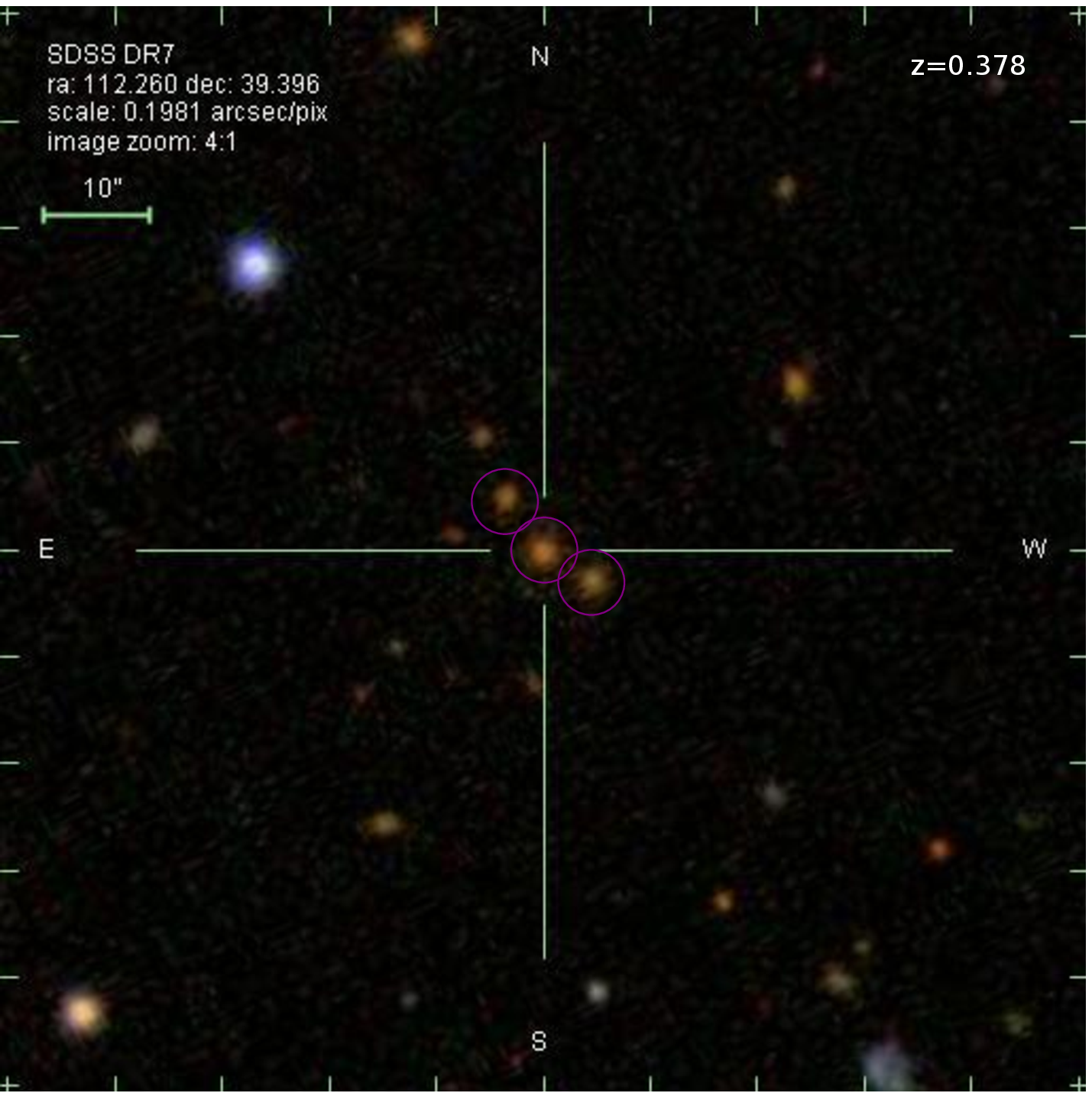}
\caption{Examples of triplets with spectroscopic and photometric measurements 
in the redshift range $0.01\le\zphot\le 0.4$. Galaxies with spectroscopy are marked with a box and triplet galaxies members with a circle. }
\label{f92}
\end{figure*}

Physical triplets are a special case of a small group with three members. 
We have selected triplet candidates comprising luminous galaxies. 
Nevertheless, from a dynamical point of view, there could be other neighbour 
galaxies with lower luminosity that affects the properties of the system.

Hereafter, we consider two proxies for galaxy 
environment: one associated to the local environment, and another 
related to larger scales. 

In order to describe global environment, we compute the distance to the closest neighbour cluster of galaxies. This quantity is related to the mass of the region 
where the galaxy resides \citep{gonzalez09}. Nevertheless, even if a 
triplet candidate system is distant from high density regions, it could still be 
part of a larger group of galaxies. To avoid such systems we have employed a fixed aperture 
method to describe the local galaxy density.

Therefore, we have selected isolated triplets that satisfies the following criteria:

\begin{enumerate}
\item $N_{05}=3$
\item $N_1 \le 4$
\item $dc\ge 5\mpc$
\end{enumerate}

$N_{05}$ and $N_1$ are the number of galaxies brighter than $M_r=-20.5$ within
$0.5\mpc$ and $1\mpc$ respectively, with the same restrictions on $\Delta V$ used to 
identify triplet members, and $dc$ is the distance to the closest neighbour cluster .

These criteria were chosen from an analysis of the triplet candidates at low redshifts ($0.01\le z<0.14$) 
where we analysed the behaviour of $N_{05}$ and $N_1$ for galaxies brighter than $M_r=-20.5$ and $M_r=-19.5$, 
in order to calibrate these two quantities for the selection of an isolation criterion to be applied at
higher redshifts where only bright galaxies are included. In order to work with a complete sample of galaxies with a fainter limiting absolute magnitude 
than that given by the spectroscopic survey, we have considered galaxies brighter than $M_r=-19.5$ up to $\zspec=0.14$, 
using both spectroscopic as well as photometric data. Accordingly, we estimate the $N_{05}$ and $N_1$ parameter taking into 
account the photometric redshift errors.

In order to analyse the global environment we used the projected distance to the closest neighbour cluster in two cluster catalogues: we used a group/cluster sample extracted from the 
catalogue constructed by \citet{zapata09} updated to the SDSS-DR7 
for $0.01 \le \zspec < 0.1$ ($d_c=d_{zapata}$). For systems with $0.1\le \zspec < 0.14$ we have used the Gaussian Mixture Brightest Cluster Galaxy (GMBCG) catalogue \citep{Hao2011} ($d_c=d_{GMBCG}$).

\citet{zapata09} employed a friends-of-friends algorithm with varying projected 
linking length $\sigma$, with $\sigma_0 = 0.239 \mpc$ and fixed radial 
linking length $\Delta V = 450 \kms$. The sample comprises 15140 groups and 
extends out to z = 0.12. Following \citet{padilla09}, we select groups with 
virial masses $M_{vir} >10^{14}h^{-1}M_{\odot}$ and define a group/cluster sample. 
The GMBCG catalogue of clusters \citep{Hao2011} comprises 55424 clusters 
identified using the red sequence plus Brightest Cluster Galaxy features, in 
the redshift range $0.1\le z<0.55$. The cluster abundance of GMBCG clusters is similar 
to \citet{zapata09} provided a mass restriction $M_{vir} >10^{14}h^{-1}M_{\odot}$ 
is applied to this sample. 
 
Therefore, for each triplet system candidate we compute the projected distance to the closest neighbour 
cluster considering a radial velocity cut $\Delta V < 1000 \kms$ when spectroscopic information 
is available and $\Delta V < 7000 \kms$ for systems with photometric redshifts.


Figure \ref{f8} shows the distribution of the distance to the closest neighbour cluster 
(left \citealt{zapata09}  and right \citealt{Hao2011}), for the spectroscopic triplet 
candidates. In black solid line the distribution for systems in the redshift range $0.01\le \zspec< 0.1$ 
(left panel) and $0.1\le \zspec<0.14$ (right panel). 
In these figures, grey dashed lines represent the distribution of triplet candidates that
satisfies $N_{05}=3$ and $N_1\le 4$, estimated using galaxies brighter
than  $M_r=-20.5$. The distribution for systems with $N_{05}=3$ and $N_1\le 4$, estimated with galaxies 
$M_r\le-19.5$ is represented in grey solid line.
In both panels of this figure it can be appreciated that the limit $dc=5\mpc$ (vertical black dashed line) encloses most of the systems with $N_{05}=3$ and $N_1\le 4$ (galaxies $M_r\le-19.5$ as well as galaxies with $M_r\le -20.5$). 

We have considered this limit in order to select a safe isolation criterion, for triplet candidates at higher redshift ($0.14\le z\le0.4$), where, by completeness, the selection of isolated triplets with $N_{05}=3$ and $N_1\le 4$ can be estimated only by using galaxies brighter than $M_r=-20.5$.

Taking a distance grater than $5\mpc$, we are ensuring that our triplets do 
not belong to any system of groups or clusters. On the other hand, by restricting 
the number of galaxies within $0.5\mpc$ and $1\mpc$ leave us with a system composed 
by only three luminous galaxies.

By imposing the isolation criterion to the sample of triplet of galaxies candidates we obtain  
$1092$ triplets in our final catalogue, $95$ triplets in the redshift range $0.01\le\zspec<0.14$ 
and $997$ in the range $0.14\le \zphot\le0.4$. 
Figure \ref{f92} shows three examples of the systems at low and intermediate redshifts. From these figures it 
can be appreciated the high degree of isolation of the systems. In a forthcoming paper 
(paper II, Duplancic et. al in preparation), we will analyse different spectro-photometric properties 
of these isolated triplet of galaxies systems.

In the Appendix we summarise some properties of 20 triplets
from the final catalogue derived in this work.
Columns 1 and 2 right ascension and declination (J2000) galaxy
positions in hexadecimal format; column 3 redshift of the central
galaxy, and column 4 number of triplet galaxy members with spectroscopic measurements
The full catalog is available electronically.


\section{Results and discussion}

In this work we have developed algorithms for the identification of galaxy triplet candidates
in the Sloan Digital Sky Survey Data Release Seven. 

Taking into account the completeness of the spectroscopic data in the range 
$0.01 \le\zspec<0.14 $ down to absolute magnitude $M_r = -20.5$, 
we identified $273$ triplet candidates using only the spectroscopic data. 

We have considered the incompleteness of these spectroscopic triplets due to fiber 
collision by using photometric as well as spectroscopic data. We find that $90\%$ 
of the spectroscopic systems can be recovered using the photometric 
data and, additionally, $39$ new triplet candidates with 1 or 2 members without spectroscopic 
information were included into this sample.

In order to explore the completeness and contamination of our procedure, 
we have employed a mock catalogue with similar volume 
than the spectroscopic S1 sample. We find a high level of completeness ($\sim 80 \%$) 
and low contamination ($\sim 5 \%$) which gives confidence to the methods 
adopted and the results obtained.

Using photometric information we have extended the sample of triplet candidates 
in the spectroscopic redshift range up to significantly larger 
depths ($0.14\le\zphot\le 0.4$). Within this larger redshift range we 
identified $5628$ triple system candidates. 

In order to build a catalogue of triplets of galaxies which fulfil the main features of physical systems, we defined an isolation criterion based on both, the distance to the closest neighbour cluster and a fixed aperture local density estimate.
This criterion ensures that the dynamics of the systems is not dominated by 
larger virialized structures and also avoids triplets contaminated by bright close neighbour galaxies.

\begin{figure*}
\begin{center}
\leavevmode \epsfysize=10cm \epsfbox{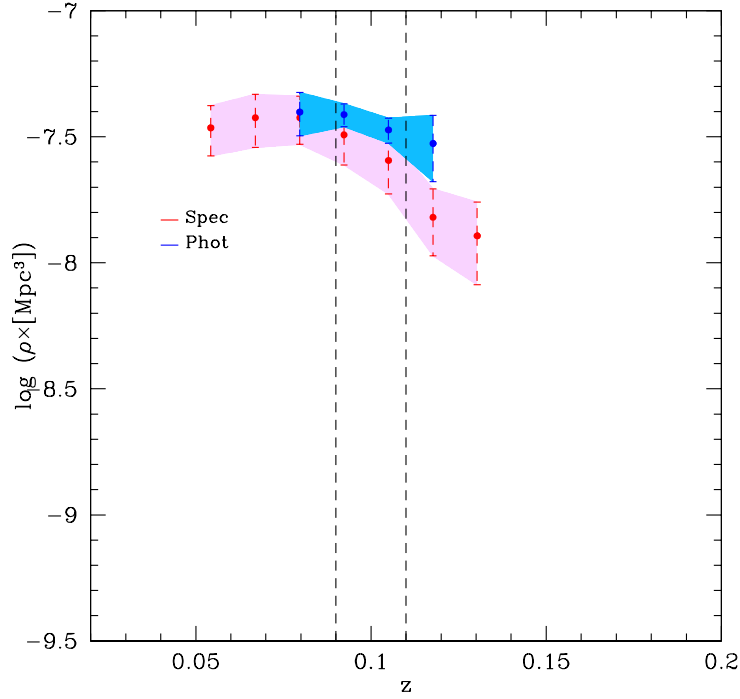}
\end{center}
\caption{Spatial number density of isolated triplet of galaxy at low redshift. Red dots represents systems 
with spectroscopic information and in blue the spatial number density with the systems incorporated by using photometric information. The error bars represent $1 \sigma$ uncertainties calculated 
using bootstrap techniques. Vertical lines shows the redshift range analysed in section 4.4.} 
\label{f10}
\end{figure*}

By imposing the isolation criterion to the sample of triplet of galaxies candidates we obtain  
$1092$ triplets in our final catalogue, $95$ triplets in the redshift range $0.01\le\zspec<0.14$ 
and $997$ in the range $0.14\le \zphot\le0.4$. 
The results can be appreciated by inspection of Figure \ref{f10}, which shows the spatial number density of isolated triplets as a function of redshift, for the low redshift sample. 
Red dots corresponds to the spectroscopic triplets and blue dots to the spatial number density with the new triplets included after the fiber collisions analysis (see section 4.4).
From this figure it can be seen that photometric redshift information is 
useful in order to complete the sample of triplets at low redshift.

\begin{figure*}
\leavevmode \epsfysize=10cm \epsfbox{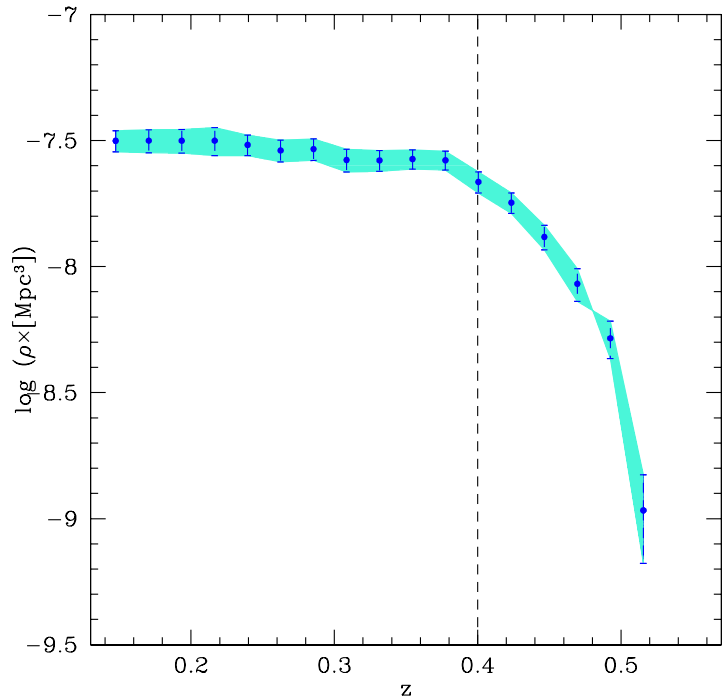}
\caption{Spatial number density of isolated triplet of galaxy at intermediate redshift.
The error bars represent $1\sigma$ uncertainties calculated using bootstrap techniques.} 
\label{f11}
\end{figure*}

Figure \ref{f11} shows the spatial number density of isolated triplets as a function of redshift for the intermediate redshift sample. By comparison to figure \ref{f10} it can be appreciated that the spatial number density of both, the intermediate and the
low redshift photometric triplet samples, smoothly match at $z\sim 0.14$. 
The observed flat behavior of the number density  up to $\zphot \sim 0.4$ shows that 
our methods are adequate to identify triplets reliably without important systematic effects.

Our results show that photometric redshifts provide very useful 
information, allowing to asses fiber collision incompleteness and extend the detection of triplets to large distances where spectroscopic redshifts are not available.

The final catalogue comprise $1092$ triplets of 
galaxies in the redshift range $0.01\le z \le0.4$. This sample is suitable for statistical analysis, which will be performed in a forthcoming paper, and may serve as targets for follow up observations.


\section{Acknowledgements}

We thank the authors Zapata, Perez, Padilla $\&$ Tissera for granting us access to their 
group catalogue. We thank the Referee for providing us with helpfull comments.
This work was supported in part by the Consejo Nacional de 
Investigaciones Cient\'ificas y T\'ecnicas de la Rep\'ublica Argentina 
(CONICET), Secretar\'\i a de Ciencia y Tecnolog\'\i a de la Universidad 
 de C\'ordoba and Secretar\'\i a de Ciencia y T\'ecnica de la Universidad Nacional
 de San Juan. Ana Laura O'Mill and Laerte Sodr\'e Jr was supported by the Brazilian 
agencies FAPESP and CNPq. 
Funding for the SDSS and SDSS-II has been provided by the Alfred P. Sloan Foundation, 
the Participating Institutions, the National Science Foundation, the U.S. 
Department of Energy, the National Aeronautics and Space Administration, 
the Japanese Monbukagakusho, the Max Planck Society, and the Higher Education 
Funding Council for England. The SDSS Web Site is http://www.sdss.org/. 
The SDSS is managed by the Astrophysical Research Consortium for 
the Participating Institutions. The Participating Institutions are 
the American Museum of Natural History, Astrophysical Institute 
Potsdam, University of Basel, University of Cambridge, Case 
Western Reserve University, University of Chicago, Drexel 
University, Fermilab, the Institute for Advanced Study, the Japan 
Participation Group, Johns Hopkins University, the Joint Institute 
for Nuclear Astrophysics, the Kavli Institute for Particle 
Astrophysics and Cosmology, the Korean Scientist Group, the 
Chinese Academy of Sciences (LAMOST), Los Alamos National 
Laboratory, the Max-Planck-Institute for Astronomy (MPIA), the 
Max-Planck-Institute for Astrophysics (MPA), New Mexico State 
University, Ohio State University, University of Pittsburgh, 
University of Portsmouth, Princeton University, the United States 
Naval Observatory, and the University of Washington.


\label{lastpage}


\appendix

\section{Catalogue summarise}
\textbf{We present a short table (table \ref{t2}) with some properties of 20 triplets
from the final catalogue derived in this work.}
\begin{table}
\caption{Examples of triplets of galaxies. Column 1 and 2: right ascension and declination 
galaxy position. Column 3: redshift of the central galaxy. Column 4: number of triplet galaxy members with spectroscopic measurements.} 
  \begin{center}\begin{tabular}{@{}ccl@{}ccl@{}cl@{}ccl@{}}
  \hline
  \hline
$\alpha$ ($J2000$) & $\delta$ ($J2000$)& redshift & & \# members with $\zspec$  \\
 \hline
07 \ 29 \ 02.43  &   39 \ 23 \ 45.76  & \   0.378    && 0 \\
07 \ 29 \ 07.89  &   45 \ 17 \ 25.93  & \   0.240    && 1 \\
09 \ 53 \ 27.44  &   49 \ 00 \ 30.45  & \   0.347    && 0 \\
10 \ 07 \ 50.36  &   00 \ 31 \ 54.62  & \   0.093    && 1 \\
10 \ 23 \ 22.62  &   08 \ 52 \ 01.11  & \   0.105    && 3 \\
10 \ 47 \ 19.52  &   31 \ 39 \ 38.56  & \   0.069    && 3 \\
12 \ 14 \ 31.24  &   08 \ 22 \ 25.71  & \   0.074    && 3 \\
12 \ 34 \ 29.58  &   32 \ 53 \ 37.44  & \   0.108    && 3 \\
12 \ 36 \ 15.72  &   05 \ 40 \ 13.24  & \   0.095    && 2 \\
13 \ 36 \ 32.90  &   25 \ 13 \ 05.15  & \   0.146    && 0 \\
13 \ 37 \ 24.89  &   41 \ 11 \ 10.15  & \   0.389    && 0 \\
14 \ 01 \ 43.37  &   21 \ 37 \ 21.52  & \   0.332    && 0 \\
14 \ 04 \ 05.64  &   08 \ 10 \ 09.85  & \   0.262    && 0 \\
14 \ 04 \ 20.47  &   08 \ 38 \ 54.09  & \   0.302    && 0 \\
14 \ 07 \ 10.10  &   64 \ 50 \ 40.58  & \   0.269    && 0 \\
14 \ 43 \ 28.41  &   13 \ 11 \ 43.88  & \   0.133    && 3 \\
15 \ 16 \ 45.33  &   07 \ 10 \ 33.34  & \   0.104    && 1 \\
15 \ 18 \ 18.93  &   05 \ 18 \ 17.95  & \   0.103    && 1 \\
15 \ 26 \ 45.25  &   18 \ 44 \ 32.07  & \   0.088    && 3 \\
16 \ 18 \ 00.75  &   25 \ 23 \ 35.13  & \   0.036    && 3 \\ 
\hline
\hline
\label{t2}
\end{tabular}
\end{center}
\end{table}



\begin{thebibliography}{99}

\bibitem[Abazajian et al.(2009)]{dr7} Abazajian, K.~N., et 
al.\ 2009, \apjs, 182, 543 
\bibitem[Aceves(2000)]{aceves} Aceves, H.\ 2000, IAU 
Colloq.~174: Small Galaxy Groups, 209, 286 
\bibitem[Agekyan \& Anosova(1967)]{aamap} Agekyan, T.~A., \& Anosova, Z.~P.\ 1967, \azh, 44, 1261 
\bibitem[Alonso et al.(2004)]{alonso04} Alonso, M.~S., Tissera, 
P.~B., Coldwell, G., \& Lambas, D.~G.\ 2004, \mnras, 352, 1081 
\bibitem[Alonso et al.(2006)]{alonso06} Alonso, M.~S., 
Lambas, D.~G., Tissera, P., \& Coldwell, G.\ 2006, \mnras, 367, 1029 
\bibitem[Anosova et al.(1992)]{anosova} Anosova, Z.~P., 
Kiseleva, L.~G., Orlov, V.~V., \& Chernin, A.~D.\ 1992, \sovast, 36, 231 
\bibitem[Blanton et al.(2003)]{blanton2003} Blanton, M.~R., Lin, 
H., Lupton, R.~H., Maley, F.~M., Young, N., Zehavi, I., 
\& Loveday, J.\ 2003, \aj, 125, 2276 
\bibitem[Blanton \& Roweis(2007)]{kcorrect} Blanton, M.~R., \& Roweis, S.\ 2007, \aj, 133, 734 
\bibitem[Chernin \& Mikkola(1991)]{chernin91} Chernin, A.~D., \& Mikkola, S.\ 1991, 
\mnras, 253, 153  
\bibitem[Costa-Duarte et al. (2011)]{mvcd11} Costa-Duarte, M. V. C., Sodr\'e, L., \& Durrett, F. \ 2011, \mnras, 411, 1716 
\bibitem[Croton et al.(2006)]{croton} Croton, D.~J., et al.\ 2006, \mnras, 365, 11 
\bibitem[Domingue \& Xu(2005)]{domingue} Domingue, D., \& Xu, C.~K.\ 2005, Bulletin of the American Astronomical Society, 37, 1455 
\bibitem[Ellison et al.(2008)]{ellison} Ellison, S.~L., Patton, 
D.~R., Simard, L., \& McConnachie, A.~W.\ 2008, \aj, 135, 1877 
\bibitem[Elyiv et al.(2009)]{elyiv09} Elyiv, A., Melnyk, O., 
\& Vavilova, I.\ 2009, \mnras, 394, 1409 
\bibitem[Fukugita et al.(1996)]{fuku96} Fukugita, M., 
Ichikawa, T., Gunn, J.~E., Doi, M., Shimasaku, K., \& Schneider, D.~P.\ 
1996, \aj, 111, 1748
\bibitem[Geller \& Huchra(1983)]{GH} Geller, M.~J., \& Huchra, J.~P.\ 1983, \apjs, 52, 61
\bibitem[Gonz{\'a}lez 
\& Padilla(2009)]{gonzalez09} Gonz{\'a}lez, R.~E., \& Padilla, N.~D.\ 2009, \mnras, 397, 1498
\bibitem[G{\'o}rski et al.(2005)]{healpix} G{\'o}rski, K.~M., 
Hivon, E., Banday, A.~J., Wandelt, B.~D., Hansen, F.~K., Reinecke, M., 
\& Bartelmann, M.\ 2005, \apj, 622, 759
\bibitem[Gunn et al.(1998)]{gunn98} Gunn, J.~E., et al.\ 1998, 
\aj, 116, 3040
\bibitem[Hao et al.(2011)]{Hao2011} Hao, J., Mckay, T., 
Koester, B., et al.\ 2011, Bulletin of the American Astronomical Society, 
43, \#213.06 

\bibitem[Hein{\"a}m{\"a}ki et al.(1998)]{heina} 
Hein{\"a}m{\"a}ki, P., Lehto, H.~J., Valtonen, M.~J., 
\& Chernin, A.~D.\ 1998, \mnras, 298, 790
\bibitem[Hern{\'a}ndez-Toledo et al.(2011)]{HT11} 
Hern{\'a}ndez-Toledo, H.~M., M{\'e}ndez-Hern{\'a}ndez, H., Aceves, H., 
\& Olgu{\'{\i}}n, L.\ 2011, \aj, 141, 74 

\bibitem[Hogg et al. (2001)]{hogg01} Hogg, D.~W., Blanton, M., 
\& SDSS Collaboration 2001, Bulletin of the American Astronomical Society, 34, 570 
\bibitem[Karachentseva(1973)]{kara73} Karachentseva, V.~E.\ 
1973, Astrofizicheskie Issledovaniia Izvestiya Spetsial'noj 
Astrofizicheskoj Observatorii, 8, 3
\bibitem[Karachentseva et al.(1979)]{karachen} Karachentseva, V.~E., Karachentsev, I.~D., 
\& Shcherbanovskii, A.~L.\ 1979, Astrofizicheskie Issledovaniia Izvestiya 
Spetsial'noj Astrofizicheskoj Observatorii, 11, 3 
\bibitem[Karachentsev \& Karachentseva(1981)]{karachen81} Karachentsev, I.~D., \& Karachentseva, 
V.~E.\ 1981, Astrofizika, 17, 5 
\bibitem[Karachentseva \& Karachentsev(1982)]{karachen82} Karachentseva, V.~E., 
\& Karachentsev, I.~D.\ 1982, Astrofizika, 18, 5 
\bibitem[Karachentseva \& Karachentsev(1983)]{karachen83} Karachentseva, V.~E., 
\& Karachentsev, I.~D.\ 1983, Astrofizika, 19, 613 
\bibitem[Karachentsev et al.(1988)]{karachen88} Karachentsev, 
V.~E., Karachentsev, I.~D., \& Lebedev, V.~S.\ 1988, Astrofizicheskie Issledovaniia 
Izvestiya Spetsial'noj Astrofizicheskoj Observatorii, 26, 42 
\bibitem[Karachentsev et al.(1990)]{karachen90} Karachentsev, 
I.~D., Karachentseva, V.~E., 
\& Lebedev, V.~S.\ 1990, NASA Conference Publication, 3098, 115
\bibitem[Karachentseva \& Karachentsev(2000)]{karachen2000} Karachentseva, V.~E., 
\& Karachentsev, I.~D.\ 2000, IAU Colloq.~174: Small Galaxy Groups, 209, 11 
\bibitem[Lambas et al.(2003)]{Lambas03} Lambas, D.~G., Tissera, 
P.~B., Alonso, M.~S., \& Coldwell, G.\ 2003, \mnras, 346, 1189 
\bibitem[Maia et al.(1989)]{Maia} Maia, M.~A.~G., da Costa, 
L.~N., \& Latham, D.~W.\ 1989, \apjs, 69, 809 
\bibitem[Merch{\'a}n \& Zandivarez(2005)]{manuel} Merch{\'a}n, M.~E., \& Zandivarez, A.\ 2005, \apj, 630, 759 
\bibitem[O'Mill et al.(2011)]{photo} O'Mill, A.~L., Duplancic, F, Garc{\'{\i}}a Lambas, D., 
\& Sodr\'e, L. \ 2010, \mnras, in press. 
\bibitem[Padilla et al.(2010)]{padilla09} Padilla, N., Lambas, 
D.~G., \& Gonz{\'a}lez, R.\ 2010, \mnras, 409, 936 
\bibitem[Pier et al.(2003)]{pier03} Pier, J.~R., Munn, J.~A., 
Hindsley, R.~B., Hennessy, G.~S., Kent, S.~M., Lupton, R.~H., \& 
Ivezi{\'c}, {\v Z}.\ 2003, \aj, 125, 1559 
\bibitem[Smith et al. (2002)]{smit02} Smith, J.~A., Tucker, 
D.~L., Allam, S.~S., \& Jorgensen, A.~M.\ 2002, Bulletin of the American 
Astronomical Society, 34, 1272 
\bibitem[\protect\citeauthoryear{Springel et al.}{2005}]{springel} Springel V., White S. D. M., Jenkins A., Frenk C. S., 
Yoshida N., Gao L., Navarro J., Thacker R., Croton D., 
Helly J., Peacock J. A., Cole S., Thomas P., Couchman 
H., Evrard A., Colberg J., Pearce F., 2005, Nature, 435, 
629
\bibitem[Strauss et al. (2002)]{mgs}
Strauss, M.~A., et al.\ 2002, \aj, 124, 1810
\bibitem[Trofimov \& Chernin(1995)]{Wide} Trofimov, A.~V., \& Chernin, A.~D.\ 1995, Astronomy Reports, 39, 308 
\bibitem[Vavilova et al.(2006)]{Vavilova} Vavilova, I.~B., Karachentseva, V.~E., Makarov, D.~I., 
\& Melnyk, O.~V.\ 2006, arXiv:astro-ph/0609622 
\bibitem[Xin-Fa et 
al.(2005)]{Xin} Xin-Fa, D., Peng, J., Jun, S., Qun, Z., \& Ji-Zhou, H.\ 2005, Astronomical and Astrophysical Transactions, 24, 187 
\bibitem[York et al.(2000)]{sdss} York, D.~G., et al.\ 2000, 
\aj, 120, 1579 
\bibitem[Zehavi et al.(2002)]{zehavi02} Zehavi, I., et al.\ 
2002, \apj, 571, 172 
\bibitem[Zapata et al.(2009)]{zapata09} Zapata, T., Perez, J., 
Padilla, N., \& Tissera, P.\ 2009, \mnras, 394, 2229 
\bibitem[Zheng et al.(1993)]{zheng93} Zheng, J.-Q., Valtonen, 
M.~J., \& Chernin, A.~D.\ 1993, \aj, 105, 2047 


\end{thebibliography}
\end{document}